\documentclass[english,reprint, superscriptaddress]{revtex4-1}
\usepackage[T1]{fontenc}
\usepackage[latin9]{inputenc}
\usepackage{amsmath}
\usepackage{amssymb}
\usepackage{graphicx}

\makeatletter
\usepackage{braket}
\usepackage{algorithm}
\usepackage{algpseudocode}

\usepackage{xcolor}
\definecolor{darkblue}{rgb}{0.1,0.2,0.6} 
\definecolor{lightblue}{rgb}{0.1,0.1,1.0}
\definecolor{darkred}{rgb}{0.8,0.1,0.2}
\usepackage[colorlinks,citecolor=lightblue,linkcolor=darkblue,urlcolor=lightblue] {hyperref}
\renewcommand{\BibitemShut}[1]{}

\makeatother

\usepackage{babel}
\begin{document}
\title{Studying dynamics in two-dimensional quantum lattices using tree tensor
network states}
\author{Benedikt Kloss}
\affiliation{Department of Chemistry, Columbia University, 3000 Broadway, New York,
New York 10027, USA}
\email{bk2576@columbia.edu}

\author{David R. Reichman}
\affiliation{Department of Chemistry, Columbia University, 3000 Broadway, New York,
New York 10027, USA}
\email{drr2103@columbia.edu}

\author{Yevgeny Bar Lev}
\affiliation{Department of Physics, Ben-Gurion University of the Negev, Beer-Sheva
84105, Israel}
\email{ybarlev@bgu.ac.il}

\date{\today}
\begin{abstract}
We analyze and discuss convergence properties of a numerically exact
algorithm tailored to study the dynamics of interacting two-dimensional
lattice systems. The method is based on the application of the time-dependent
variational principle in a manifold of binary and quaternary Tree
Tensor Network States. The approach is found to be competitive with
existing matrix product state approaches. We discuss issues related
to the convergence of the method, which could be relevant to a broader
set of numerical techniques used for the study of two-dimensional
systems.
\end{abstract}
\maketitle

\section{\label{sec:Introduction}Introduction}

The exact simulation of the non-equilibrium dynamics of interacting
quantum lattice systems is generally an unsolved challenge, due to
the exponential growth of the Hilbert space with the size of the system.
Tensor network state (TNS) methods allow for a significant extension
of accessible length scales by trading in the exponential cost in
system size for an exponential cost in time. This becomes possible
due to a reduction of the exact Hilbert space in terms of a structured
product of low-order tensors, referred to as a tensor network. The
set of the states expressible by a given tensor network spans only
a small region in the full Hilbert space, but the coverage can be
improved systematically by increasing the number of variational parameters,
i.e. the bond dimension. For partitions of the lattice that lead to
simply-linked tensor network parts, the logarithm of the bond-dimension
gives an upper bound to the entanglement entropy. Since the entanglement
of a generic system after a global quench grows linearly with time
\citep{Calabrese_2005,chiara2006entanglement,Alba7947}, the accessible
timescales are limited. In one-dimensional systems, these timescales
are often comparable to those attainable in experimental realizations
\citep{RevModPhys.80.885}, however going to higher spatial dimensions
becomes extremely challenging due to a number of reasons.

\begin{figure}
\includegraphics[width=1\columnwidth]{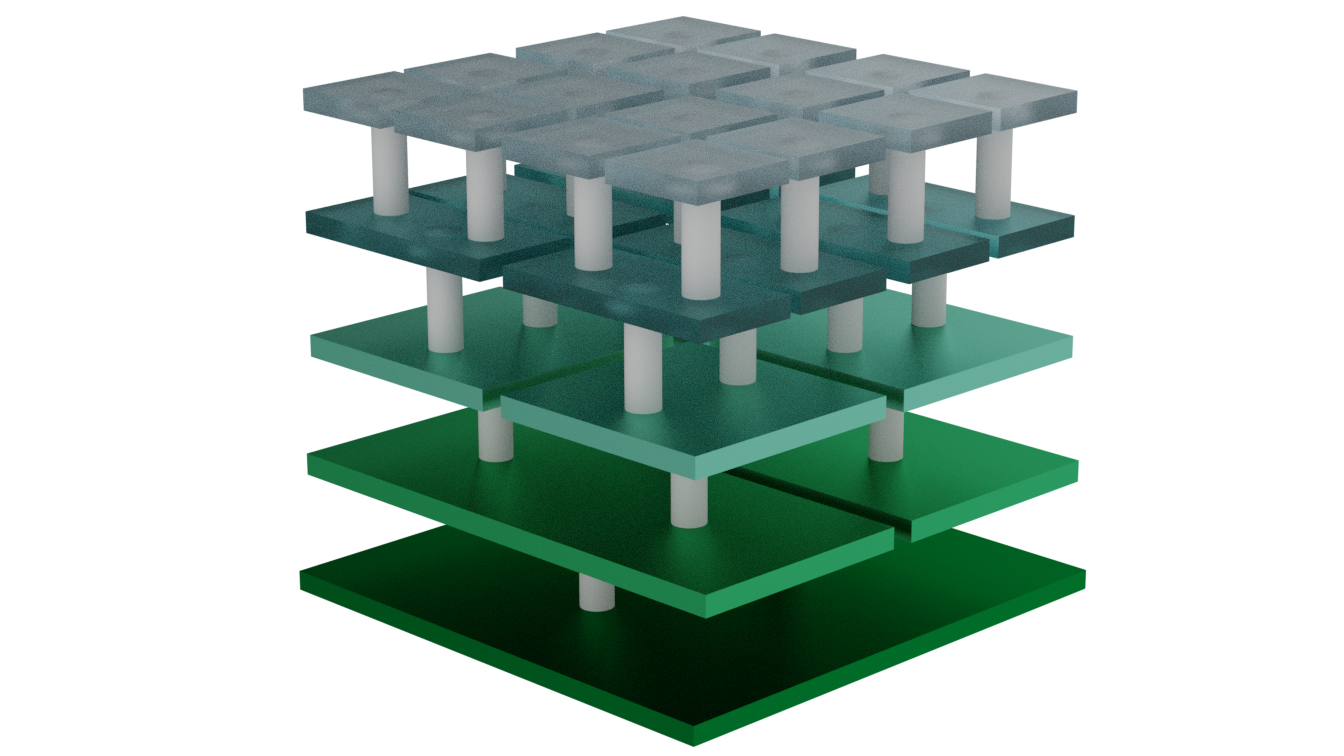}

\caption{\label{fig:binary}Illustration of binary TTNS structure for a 4x4
lattice. The physical degrees of freedom are on the topmost layer
and the top node is in the bottom-layer of the figure.}
\end{figure}
While in one-dimensional systems, matrix product states (MPS) are
known to efficiently represent area-law entangled states (which includes
ground-states of gapped one-dimensional systems), this does not hold
in two spatial dimensions \citep{PhysRevLett.69.2863,Hastings_2007,SCHOLLWOCK201196}.
The generalization of MPS to two-dimensional lattices is called Projector-Entangled
Pair States (PEPS) \citep{verstraete2004renormalization}, which provides
an efficient representation of two-dimensional area-law entangled
states \citep{verstraete2006criticality}, but PEPS are challenging
to manipulate numerically \citep{lubasch2014unifying} (see also Ref.~\citealp{ran2017lecture}
for a recent review). Approximations that are hard to control are
typically used in PEPS algorithms in order to tame the computational
effort. Even with such approximations, the computational scaling is
usually unfavorably steep. Nonetheless, PEPS-derived methods are state-of-the
art numerical techniques for computing ground-states of two-dimensional
systems \citep{zheng2017stripe}. Extensions of PEPS methods to the
time-domain have been recently developed, however the accessible timescales
are extremely limited~\citep{pivzorn2011time,kshetrimayum2019time,hubig2019evaluation,zaletel2019isometric}.
An alternative approach is to use tensor network structures, which
are more numerically tractable. One way to achieve this is to map
the two-dimensional lattice into a one-dimensional chain and apply
MPS methods, which are adjusted to handle the long-ranged interactions
that arise from the mapping \citep{crosswhite2008applying,frowis2010tensor,stoudenmire2012studying,Zaletel2015,paeckel2019time,doggen2020slow}.
Ref.~\citealp{Zaletel2015}, for example, introduced an algorithm
which expresses the propagator as a matrix product operator (MPO)
acting on the states encoded as MPS. The application of this approach
to two-dimensional lattices shares the very limited timescale of the
more recent approaches based on PEPS, since the advantages in the
computational scaling of simpler tensor networks are balancing out
the disadvantages in non-optimal representation of entanglement by
the tensor network structure for the problem at hand. A novel development
is the use of artificial neural networks (ANN) to encode the wavefunction
and its time-evolution~\citep{carleo2017solving}. They have been
shown to perform competitively with state-of-the art TNS techniques
in recent applications to two-dimensional systems~\citep{schmitt2019quantum,lpezgutirrez2019real}.
However, much remains to be learned about the possibilities and limitations
of such methods.

\begin{figure}
\includegraphics[width=1\columnwidth]{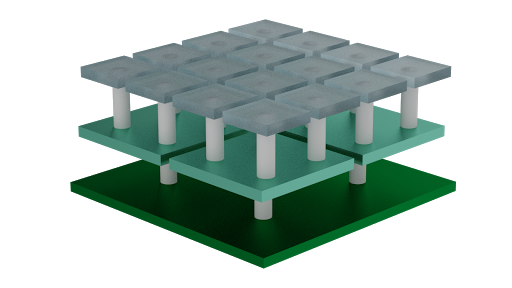}

\caption{\label{fig:quaternary}Same as Fig.~\ref{fig:binary} but for a quaternary
TTNS.}
\end{figure}
It is important to explore computationally tractable tensor network
structures other than MPS, since they may enable progress in the computation
of the exact dynamics of interacting two-dimensional systems. For
this purpose, in this work we propose to employ Tree Tensor Network
States (TTNS), which encompass all loop-free tensor network states.
While similarly to MPS, hierarchical, tree-like TTNS can only efficiently
encode states with area-law entanglement in one dimensional systems
they offer a more robust description of ground states of \emph{critical}
one dimensional systems \citep{Rizzi_2010,tsai2020tensor}, and therefore
might provide more flexibility in encoding complex entanglement structures
in two and higher dimensional systems. TTNS are used in the context
of interacting lattice systems~\citep{PhysRevA.74.022320,Tagliacozzo2009,PhysRevB.82.205105,PhysRevB.86.195137,Gerster2014,PhysRevB.96.195123,chepiga2019comb,milsted2019tensornetwork,PhysRevResearch.2.013145},
but they feature more prominently in applications like electronic
structure methods \citep{Nakatani2013,Murg2015,gunst2018t3ns} or
molecular quantum dynamics in the chemical physics literature. In
this context they are called the Multi-Configuration Time-Dependent
Hartree (MCTDH) method and its multi-layer generalization (ML-MCTDH)
\citep{meyer1990multi,beck2000multiconfiguration,wang2003multilayer}.
In ML-MCTDH, the time-dependent variational principle \citep{dirac_1930,frenkel1934wave}
(TDVP) is applied to a TTNS as a variational ansatz for the wavefunction.
Up to differences in the numerical integrations scheme, these methods
are similar to the more recent applications of the TDVP tailored specifically
to matrix product states \citep{PhysRevLett.107.070601,PhysRevLett.109.267203,PhysRevLett.111.207202,Lubich2014,Haegeman2016}.

The TDVP applied to TNS has been discussed as a method that may enable
the accurate description of hydrodynamic transport in non-integrable
systems when used with a moderate bond dimension \citep{leviatan2017quantum},
but was shown to not be a robust approximation for generic systems
\citep{PhysRevB.97.024307}. Several tensor network techniques have
been designed to circumvent the entanglement growth on intermediate
timescales with the goal of a reliable approximation to the long-time
dynamics \citep{PhysRevLett.117.210402,krumnow2019overcoming,PhysRevLett.124.137701,PhysRevB.98.235163,PhysRevB.99.235115,PhysRevB.97.035127,rakovszky2020dissipationassisted}.
Despite promising results, the stability of such approximations for
generic systems, especially beyond one dimension, is not sufficiently
established at this point. In this work, we thus consider the TDVP
applied to TNS as a numerically exact technique, allowing to compute
the dynamics of a system within a controllable accuracy up to some
finite time, and generalize the algorithms of Refs.~\citealp{Lubich2014a,Lubich2014,Haegeman2016}
to general TTNS. We note in passing that such algorithms have been
used to find the ground state of a two-dimensional spin system \citep{PhysRevB.87.125139}
and to obtain the dynamics of a zero-dimensional model \citep{schroder2019tensor}.
Recently, similar versions of this algorithm were reported in detail
in Refs.~\citealp{bauernfeind2019time,ceruti2020time}, which we
became aware of during the preparation of this manuscript. While in
our work we focus on two-dimensional systems, Ref.~\citealp{bauernfeind2019time}
showcases a promising application of a TTNS as an impurity-solver,
which is an effectively zero-dimensional problem. On the other hand,
Ref.~\citealp{ceruti2020time} proves the algorithm's exactness property
as well as a linear error-bound for the total time evolution in the
time-step.

The purpose of this work is to investigate the performance of TTNS
as a numerically exact method to study the dynamics of two-dimensional
systems. In Sec. \ref{sec:Theory}, we introduce the main concepts
of TTNS along with the TDVP before presenting the algorithm and commenting
on some caveats which are relevant to the applications of the TDVP.
We benchmark the method on an exactly solvable, non-interacting two-dimensional
system, and compare our approach to previously published results for
two-dimensional interacting hardcore bosons in Sec.~\ref{sec:Results}.
Notably, we identify the reachable timescales and investigate convergence
properties of the algorithm alongside with practical considerations
regarding how to assess the accuracy of the results. We conclude by
placing the results in the context of existing techniques and recent
developments in Sec.~\ref{sec:Conclusion}.

\section{\label{sec:Theory}Theory}

\begin{figure}
\includegraphics[width=1\columnwidth]{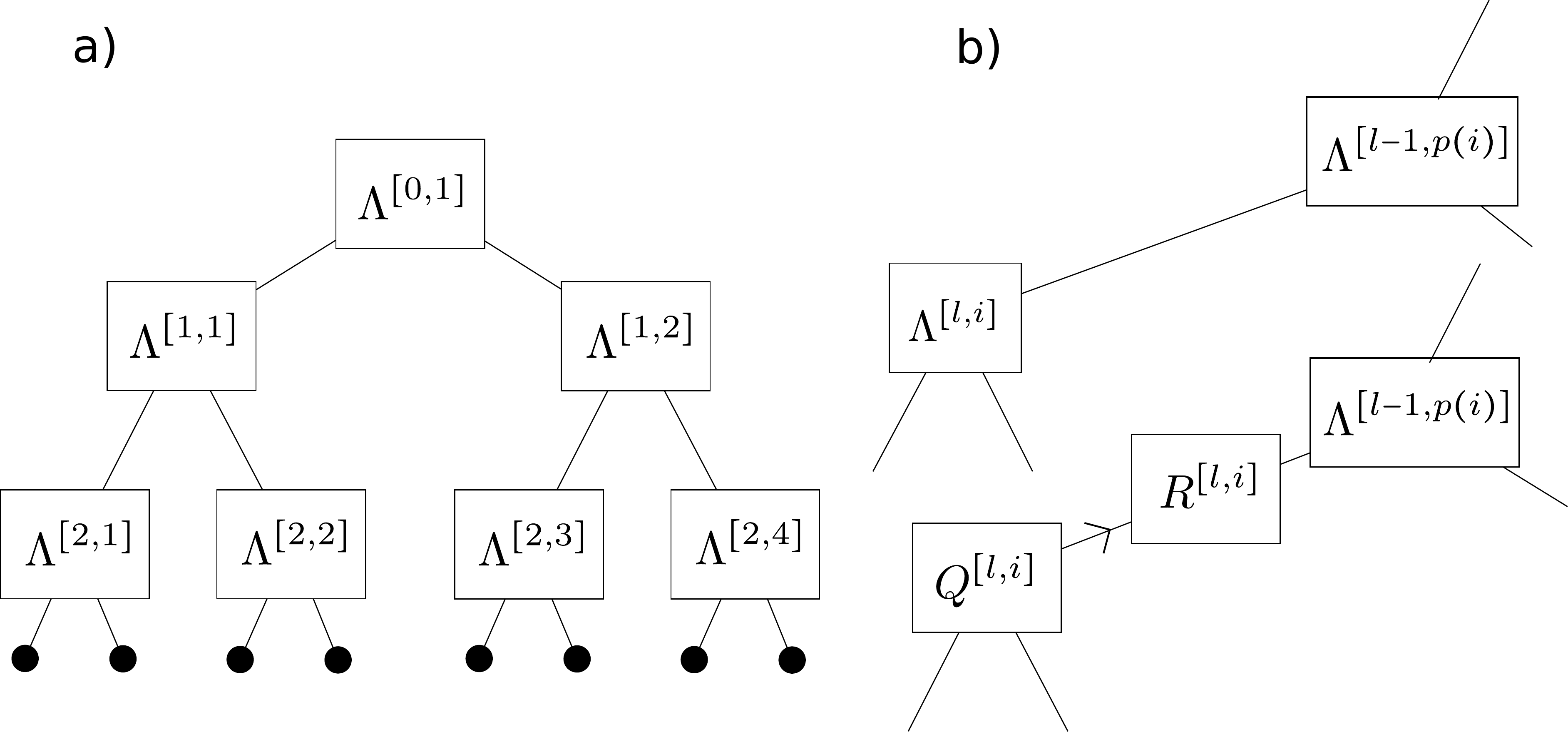}

\caption{\label{fig:TTNS_uniso}a) A binary TTNS for an 8-site system. The
black dots correspond to physical sites and the square boxes with
$n$ legs represent $n$-th order tensors. b) Application of the QR
decomposition to tensors in the TTNS. The upper and lower diagram
represent the same TTNS. The arrow on the link indicates the direction
along which the tensor is orthonormalized.}
\end{figure}

Tensor network states represent a pure state, $\ket{\Psi}=\sum_{s_{1}\dots s_{N}}\Psi_{s_{1}\dots s_{N}}\ket{s_{1}\dots s_{N}}$,
of a lattice system as a product of tensors $\{T\}.$ Each tensor
$T_{i}$ may have a number of indices corresponding to physical degrees
of freedom and also auxiliary indices which do not correspond to physical
degrees of freedom. Consider the Schmidt decomposition, corresponding
to a bipartition of the lattice into a set of sites $A$ and its complement
$B$, $\Psi_{s_{1}\dots s_{N}}=\sum_{i,j}\phi_{i\mathbf{s_{A}}}^{A}\lambda_{ij}\phi_{j\mathbf{s_{B}}}^{B}$
with $\lambda_{ij}=\delta_{ij}\lambda_{i}$. This expression can be
also understood as a product of three tensors, where a single auxiliary
index of each tensor is shared with the diagonal matrix of the Schmidt
coefficients (or singular values) $\lambda_{i}$. In a general tensor
network any auxiliary index will appear on two tensors, and summation
over the common index implies contraction of the two tensors. Tensor
networks can be represented diagrammatically, see Fig.~\ref{fig:TTNS_uniso}a,
where the nodes correspond to tensors and the links, dubbed \emph{legs}
in the following, indicate a shared index between the two tensors.
Any tensor network for which the legs do not form closed loops is
considered a \emph{Tree Tensor Network (TTN)}, with matrix product
states (MPS) serving as a prominent special case, which is mostly
applicable for one-dimensional lattices. Here, we focus on more general
TTNS with a simple hierarchical structure: \emph{n-ary TTNS} in which
every node has one \emph{parent} node and $n$ \emph{child} nodes,
except for those in the top and bottom layers. We group all physical
degrees of freedom into the bottom layer such that all layers above
the bottom layer contain only nodes with auxiliary legs (see Figs.\,\ref{fig:TTNS_uniso}a)--\ref{fig:quaternary}
for illustration). Without restricting the generality, in this Section
we will limit the discussion to binary TTNS. In such TTNS, a general
node represents a third order tensor $\Lambda^{[l,i]}$, where $l$
denotes the layer of the tree to which the node belongs, and $i$
enumerates the nodes in that layer. Each such node has two child nodes.
Due to the lack of loops in the tensor network, the physical degrees
of freedom separate naturally into two groups from the perspective
of a node $\Lambda^{[l,i]}$: those reachable by only descending in
the tree towards the bottom layer, i.e. those in the subtree of $\Lambda^{[l,i]}$,
and their complement. We define the number of non-zero singular values
of the Schmidt decomposition along this bipartition as the rank $r$
of node $\Lambda^{[l,i]}$. For a state with volume law entanglement,
the exact rank $r$ will generally scale exponentially with the system
size. Thus we introduce a cutoff in the number of kept singular values,
namely the bond dimension of the tree $\chi$. In the following, we
consider a TTNS of rank $\chi$, which implies that all its tensors
$\Lambda^{[l,i]}$ have a rank of $\min\left(\chi,d^{N(l,i)}\right)$,
where $d$ denotes the local Hilbert space dimension and $N(l,i)$
is the number of sites in the subtree of $\Lambda^{[l,i]}$. The set
of TTNS with a given rank $\mathbf{\chi}$ constitutes a smooth manifold
of states $\mathcal{M_{\mathbf{\chi}}}$. The computational complexity
for a binary TTNS is $\mathcal{O}\left(N\log N\chi^{3}\right)$ in
memory and $\mathcal{O}\left(N\log N\chi^{4}\right)$ in computation
where $N$ is the number of physical degrees of freedom.

We next present a method for time-propagation on the manifold $\mathcal{M}_{\chi}$
of tree tensor networks with tree rank $\chi$ using a time-dependent
variational principle (TDVP) \citep{dirac_1930,frenkel1934wave}.
We start by introducing a few properties and manipulations of TTNS
and then describe TDVP and its application to TTNS. We finally highlight
important technical details in the use of the TDVP.

\subsection{TTNS - Basics}

A TTNS of a rank $\chi$ is unique up to unitary transformations.
This can be seen by inserting a unit matrix between two linked nodes
of the tree
\begin{multline}
\Lambda_{\alpha_{1}\alpha_{2}\alpha_{3}}^{[l,i]}\mathbb{\mathbb{I}}_{\alpha_{3}\beta_{1}}\Lambda_{\beta_{1}\beta_{2}\beta_{3}}^{[l+1,j]}=\Lambda_{\alpha_{1}\alpha_{2}\alpha_{3}}^{[l,i]}U_{\alpha_{3}\gamma}^{*}U_{\gamma\beta_{1}}\Lambda_{\beta_{1}\beta_{2}\beta_{3}}^{[l+1,j]}\\
=\tilde{\Lambda}_{\alpha_{1}\alpha_{2}\gamma}^{[l,i]}\tilde{\Lambda}_{\gamma\beta_{2}\beta_{3}}^{[l+1,j]},\label{eq:TTNS_unitary}
\end{multline}
where repeated indices are summed over, $\mathbb{\mathbb{I}}$ represents
a $\chi\times\chi$ unit matrix and $U^{*}$ indicates complex conjugation
of the corresponding tensor. This property can be exploited to \emph{isometrize}
the tree around any of its nodes \citep{PhysRevB.77.214431,Gerster2014},
which is a generalization of the mixed canonical representation of
MPS. To illustrate this concept, consider the \emph{isometrization}
about the top-node. In this perspective, every tensor in the tree,
except the top-node, represents a truncated orthonormal basis in the
space of the bases of child nodes, called \textit{isometry} in the
language of real-space or tensor RG. Through recursion, a structured,
incomplete basis for the physical lattice sites is obtained. The coefficients
for these basis functions are contained in the top node. Any general
TTNS can be brought into this form using a sequence of QR decompositions.
Practically, one applies QR factorization $\Lambda_{\alpha_{1}\alpha_{2}\alpha_{3}}^{[l,i]}=Q_{\beta\alpha_{2}\alpha_{3}}^{[l,i]}R_{\beta\alpha_{1}}^{[l,i]}$
with $Q_{\beta\alpha_{2}\alpha_{3}}^{[l,i]*}Q_{\gamma\alpha_{2}\alpha_{3}}^{[l,i]}=\delta_{\beta,\gamma}$,
for each of the nodes proceeding layer by layer from bottom to top
and absorbing the matrices $R$ into the parent node after each factorization
(see also Fig.\  \ref{fig:TTNS_uniso}b).
\begin{figure}
\includegraphics[width=1\columnwidth]{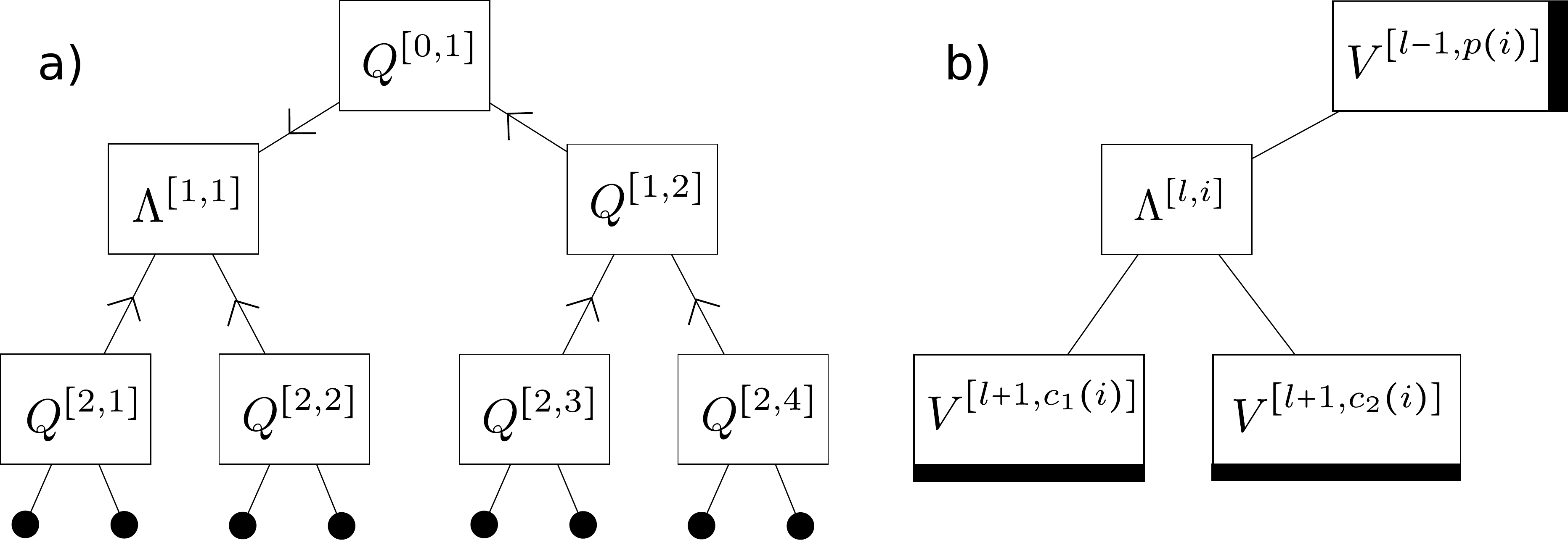}

\caption{\label{fig:TTNSiso}A TTNS isometrized about node {[}1,1{]}, a), and
its shorthand notation, b). The thick black bars on the environment
tensors represent the set of physical sites belonging to each of the
environment tensors. Note that orthogonality of the environment tensors
in b) is not indicated by arrows on the legs, but implicit in their
definition.}
\end{figure}
Graphically, the direction along which the tensors are orthogonalized
is indicated by an arrow on the linking leg. Isometrization around
a specific node in the tree translates into arrows pointing in the
direction of this node on any (direct) path between the node and a
physical site, see Fig.\,\ref{fig:TTNSiso}b). We may rewrite the
TTNS in the following manner:
\begin{equation}
\Psi[l,i]_{\mathbf{s}}=\Lambda_{\alpha_{1}\alpha_{2}\alpha_{3}}^{[l,i]}V_{\alpha_{1}\mathbf{s}_{1}}^{[l-1,p(i)]}V_{\alpha_{2}\mathbf{s}_{2}}^{[l+1,c_{1}(i)]}V_{\alpha_{3}\mathbf{s}_{3}}^{[l+1,c_{2}(i)]}.\label{eq:Isom_env}
\end{equation}
Here, we take the TTNS to be isometrized about node $[l,i]$, indicated
as $\Psi[l,i]$, and an environment tensor $V_{\alpha_{j}\mathbf{s}_{j}}^{[l\pm1,p(i)/c_{j}(i)]}$
is the contraction of all tensors between the legs of node $\Lambda^{[l,i]}$,
labeled by $\alpha_{j}$, and the physical sites $\mathbf{s}_{j}$,
linked by paths from leg $\alpha_{j}$ that do not cross node $\Lambda^{[l,i]}$.
$c_{j}(i)$ and $p(i)$ are placeholders for the child and parent
of node $\Lambda^{[l,i]}$, respectively. We note in passing that
similarly to MPS methods, such a contraction is never explicitly carried
out, and we only use it for notational convenience. For future reference,
we define projectors onto environment tensors of the lower and upper
levels in the hierarchy: $\left(\Omega^{[l+1,c_{j-1}(i)]}\right){}_{\mathbf{s}'_{j}\mathbf{s}_{j}}=V_{\alpha_{j}\mathbf{s'}_{j}}^{[l+1/c_{j-1}(i)]}V_{\alpha_{j}\mathbf{s}_{j}}^{[l+1/c_{j-1}(i)]*}$
and $\left(\Omega^{[l-1,p(i)]}\right){}_{\mathbf{s}'_{1}\mathbf{s}_{1}}=V_{\alpha_{1}\mathbf{s'}_{1}}^{[l-1/p(i)]}V_{\alpha_{1}\mathbf{s}_{1}}^{[l-1/p(i)]*}$.
A useful property of the environment tensors is their orthogonality,
which allows for efficient calculation of certain physical quantities.
For example, if the state is isometrized about node $[l,i]$, the
norm of the state is given by $\braket{\Psi[l,i]|\Psi[l,i]}=\Lambda_{\alpha_{1}\alpha_{2}\alpha_{3}}^{[l,i]*}\Lambda_{\alpha_{1}\alpha_{2}\alpha_{3}}^{[l,i]}$
since $V_{\alpha_{j}\mathbf{s}_{j}}^{[l\pm1,p(i)/c_{j}(i)]*}V_{\alpha'_{j}\mathbf{s}_{j}}^{[l\pm1,p(i)/c_{j}(i)]}=\delta_{\alpha_{j}'\alpha_{j}}$
. To improve the readability of the presentation, in the following
we will omit the indices specifying the elements of the tensors.

\subsection{TDVP}

\begin{figure}
\includegraphics[width=0.95\columnwidth]{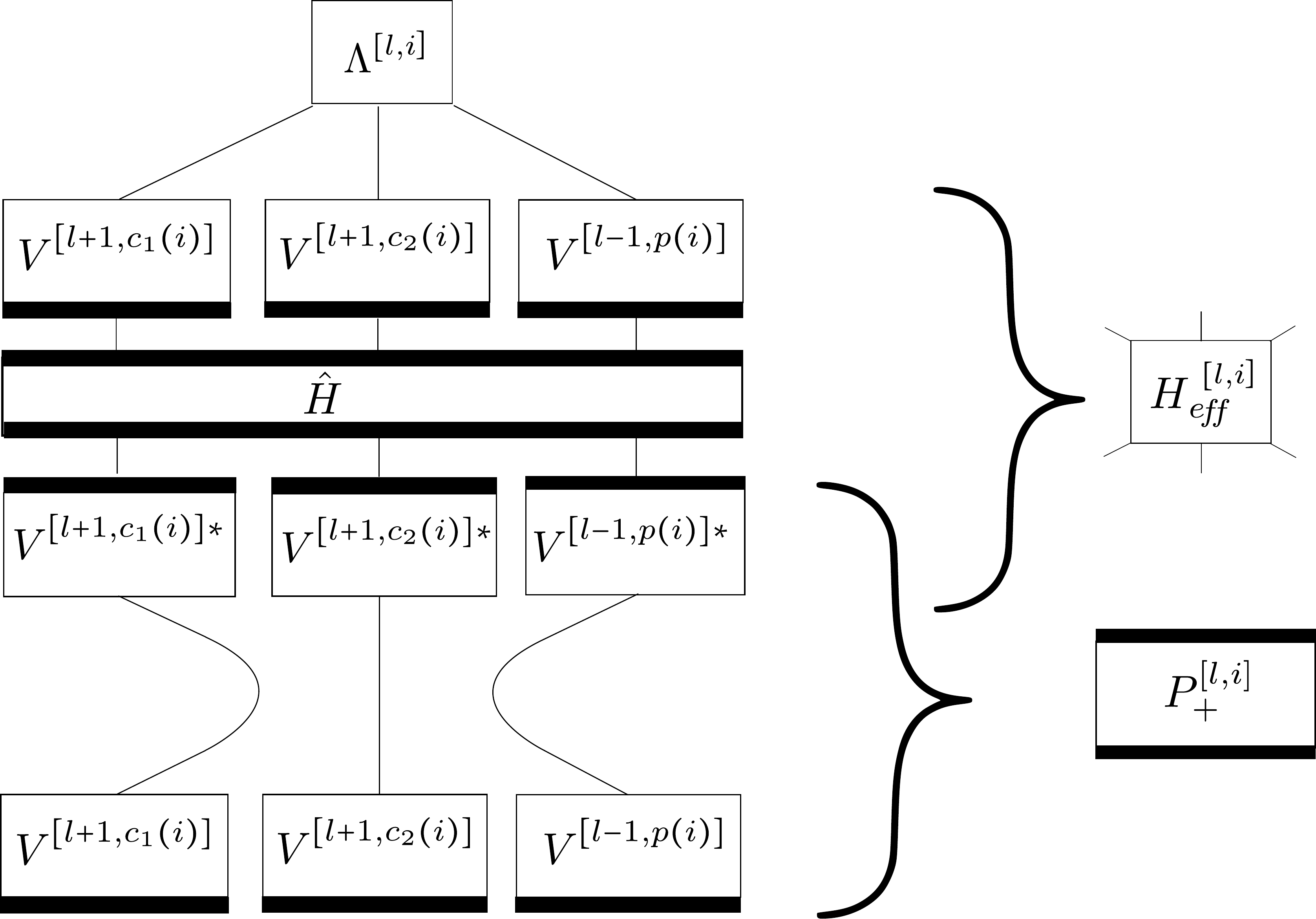}

\caption{\label{fig:Heff}Graphical representation of the last line of Eq.\ (\ref{eq:idtPsi_P+}),
with identification of the effective Hamiltonian environment, $H_{eff}^{[l,i]}$,
of Eq.\ (\ref{fig:Heff}) as well as part of the tangent space projector,
$P_{+}^{[l,i]},$ of Eq.\ (\ref{eq:PTang_P+}). In contrast to Fig.\ \ref{fig:TTNSiso},
the environment tensors have been brought on the same level regardless
of layer for better readability.}
\end{figure}

The time-dependent variational principle generates classical dynamics
in the space of variational parameters, $\alpha$, described by the
Lagrangian
\begin{equation}
\mathcal{L}[\alpha,\dot{{\alpha}}]=\bra{\Psi[\alpha]}i\partial_{t}\ket{\Psi[\alpha]}-\bra{\Psi[\alpha]}\hat{H}\ket{\Psi[\alpha]}.\label{eq:Lagrangian}
\end{equation}
The associated action is minimized along a path on a certain variational
manifold, which in our case is the manifold of TTNS with tree rank
$\chi$, $\mathcal{M}_{\chi}$. The principle of least-action yields
the following equation of motion, 
\begin{equation}
i\partial_{t}\ket{\Psi[\alpha]}=P_{T}(\Psi[\alpha])\hat{H}\ket{\Psi[\alpha]},\label{eq:TDPSI_PT}
\end{equation}
where $P_{T}(\Psi[\alpha])$ is the projector onto the tangent space
of the manifold $\mathcal{M}_{\chi}$ at the point $\Psi[\alpha]$.
An expression for $P_{T}(\Psi[\alpha])$ was derived for general binary
TTNS in Refs. \citealp{uschmajew2013geometry,lubich2013dynamical}.
Here, we will use an additive splitting of $P_{T}(\Psi[\alpha])$,
in an analogy to those presented for TTNS with only two layers, i.e.
Tucker tensors \citep{Lubich2014a} and matrix product states \citep{Lubich2014,Haegeman2016},
respectively. Note that the latter two TNS are subclasses of a general
TTNS and that the expressions for the projector, $P_{T}(\Psi[\alpha])$,
is not restricted to binary TTNS and is valid for \emph{any} TTNS
with straightforward modifications. In particular, 
\begin{equation}
P_{T}(\Psi[\alpha])=P_{0}+\sum_{[l,i]}P_{+}^{[l,i]}-P_{-}^{[l,i]}\label{eq:PTang_full}
\end{equation}
with
\begin{eqnarray}
P_{0} & = & \Omega^{[1,1]}\Omega^{[1,2]}\label{eq:PTang_P0}\\
P_{+}^{[l,i]} & = & \Omega^{[l+1,c_{1}(i)]}\Omega^{[l+1,c_{2}(i)]}\Omega^{[l-1,p(i)]}\label{eq:PTang_P+}\\
P_{-}^{[l,i]} & = & \Omega^{[l,i]}\Omega^{[l-1,p(i)]}.\label{eq:PTang_P-}
\end{eqnarray}
Inserting this splitting into (\ref{eq:TDPSI_PT}) leads to a set
of projected Schrödinger equations for the tensors $\Lambda^{[l,i]}$
and matrices $R^{[l,i]}$. For example under the action of $P_{+}^{[l,i]}$
(see also \,\ref{fig:Heff}):
\begin{multline}
i\partial_{t}\Psi[\alpha]=i\dot{\Lambda}^{[l,i]}(V^{[l-1,p(i)]}V^{[l+1,c_{1}(i)]}V^{[l+1,c_{2}(i)]})+\\
i\Lambda^{[l,i]}\partial_{t}(V^{[l-1,p(i)]}V^{[l+1,c_{1}(i)]}V^{[l+1,c_{2}(i)]})\\
=(V^{[l-1,p(i)]}V^{[l+1,c_{1}(i)]}V^{[l+1,c_{2}(i)]})\\
H_{eff}^{[l,i]}\Lambda^{[l,i]},\label{eq:idtPsi_P+}
\end{multline}
with the effective Hamiltonian environment
\begin{multline}
H_{eff}^{[l,i]}=\\
(V^{[l-1,p(i)]}V^{[l+1,c_{1}(i)]}V^{[l+1,c_{2}(i)]})^{*}\hat{H}\\
(V^{[l-1,p(i)]}V^{[l+1,c_{1}(i)]}V^{[l+1,c_{2}(i)]}).\label{eq:Henv}
\end{multline}
We choose a convenient gauge in which the time-derivative of any tensor
of the TTNS representation is orthogonal to itself. This must be done
to avoid over-completeness of the basis of the tangent space. In this
gauge the time derivative simplifies to
\begin{equation}
i\dot{\Lambda}^{[l,i]}=H_{eff}^{[l,i]}\Lambda^{[l,i]},\label{eq:LambdaDot}
\end{equation}
which is obtained by contracting Eq.\,(\ref{eq:idtPsi_P+}) with
$(V^{[l-1,p(i)]}V^{[l+1,c_{1}(i)]}V^{[l+1,c_{2}(i)]})^{*}$. Similarly,
we obtain for $R^{[l,i]}$, which results from the action of $P_{-}^{[l,i]}$,

\begin{equation}
i\dot{R}^{[l,i]}=\tilde{H}_{eff}^{[l,i]}R^{[l,i]}.\label{eq:Rdot}
\end{equation}
with the effective Hamiltonian environment
\begin{equation}
\tilde{H}_{eff}^{[l,i]}=(V^{[l-1,p(i)]}V^{[l,i]})^{*}\hat{H}(V^{[l-1,p(i)]}V^{[l,i]})\label{eq:Heff_forR}
\end{equation}
Time-evolution is obtained by integrating the linear differential
equations Eqs.~(\ref{eq:LambdaDot}) and (\ref{eq:Rdot}) using the
projector splitting integrator. Evaluating the action of the Hamiltonian
environments in Eqs. (\ref{eq:Henv}) and (\ref{eq:Heff_forR}) generally
requires a compressed representation of the Hamiltonian, for example
as a matrix product operator (MPO) or tree tensor network operator
(TTNO), in which case the environments are recursively contractible
with the TTN. Alternatively one can express the Hamiltonian as a sum
of rank-1 terms, in which case evaluating Eqs.~(\ref{eq:LambdaDot})
and (\ref{eq:Rdot}) simplifies to a sum over matrix multiplications
applied to the tensor for which the time-derivative is calculated.
The number of Hamiltonian terms to be evaluated for a given site can
be reduced by combining terms in the rank-1 decomposition of the Hamiltonian
during the recursive contraction.

\subsection{Splitting integrator}

Formally, the splitting integrator is obtained using a Trotter splitting
applied to the additive decomposition of the tangent-space-projected
evolution operator. Practically, it consists of a forward walk on
the tree, propagation of the top-level tensor $\Lambda^{[0,1]}$ for
a full time step, and a backward walk on the tree. A pseudo-code is
given in algorithms \ref{alg:forward}-\ref{alg:core}. During the
walks on the tree, isometrization of the TTNS is always maintained
about the currently visited node and the effective Hamiltonian matrices
are updated when going from one node to another along the direction
of the step. The forward walk (backward walk) starts from the top-level
node and proceeds from the current node to the adjacent node in a
clockwise direction (in a counter-clockwise direction) closest to
the previous/incoming node and propagation for half a time step is
performed only while ascending (descending). A walk on the tree is
finished once the top-node is reached after visiting all physical
sites, i.e. after each tensor (and the associated matrix $R$) is
propagated save those of the top node.

\begin{figure} \begin{algorithm}[H]
\caption{Forward walk}
\label{alg:forward}
\begin{algorithmic}[1]
\Require $\Psi[l,i]$, current node $[l,i]$, next node $[l-1,p(i)]$
\Ensure $\Psi[l-1,p(i)]$
\If {in forward loop}:
\State $\Lambda^{[l,i]}(t_{1/2}) \leftarrow \mathrm{propagate}(\Lambda^{[l,i]}(t_0),h/2)$
\State compute QR fact. $\Lambda^{[l,i]}(t_{1/2})= Q^{[l,i]}(t_{1/2})R^{[l,i]}(t_{1/2})$
\State $\Lambda^{[l,i]}(t_{1/2}) \leftarrow Q^{[l,i]}(t_{1/2})$
\State $R^{[l,i]}(t_{0}) \leftarrow \mathrm{propagate}(R^{[l,i]}(t_{1/2}),-h/2)$
\State $\Lambda^{[l-1,p(i)]}(t_0) \leftarrow \leftarrow Q^{[l-1,p(i)]}(t_{0})R^{[l,i]}(t_{0})$
\Else
\State compute QR fact. $\Lambda^{[l,i]}(t_{1})=Q^{[l,i]}(t_{1})R^{[l,i]}(t_{1})$
\State $\Lambda^{[l,i]}(t_1) \leftarrow Q^{[l,i]}(t_{1})$
\State $\Lambda^{[l-1,p(i)]}(t_1) \leftarrow Q^{[l-1,p(i)]}(t_{1})R^{[l,i]}(t_{1})$

\EndIf
\end{algorithmic} \end{algorithm}

\begin{algorithm}[H]
\caption{Backward walk}
\label{alg:backward}
\begin{algorithmic}[1]
\Require $\Psi[l,i]$, current node $[l,i]$, next node $[l+1,c_{j}(i)]$
\Ensure $\Psi[l+1,c_{j}(i)]$
\If {in backward loop}:
\State compute QR fact. $\Lambda^{[l,i]}(t_{1})= Q^{[l,i]}(t_{1})R^{[l,i]}(t_{1})$
\State $\Lambda^{[l,i]}(t_{1}) \leftarrow Q^{[l,i]}(t_{1})$
\State $R^{[l,i]}(t_{1/2}) \leftarrow \mathrm{propagate}(R^{[l,i]}(t_{1}),-h/2)$
\State $\Lambda^{[l+1,c_j(i)]}(t_{1/2}) \leftarrow Q^{[l+1,c_j(i)]}(t_{1/2})R^{[l,i]}(t_{1/2})$
\State $\Lambda^{[l+1,c_j(i)]}(t_{1}) \leftarrow \mathrm{propagate}(\Lambda^{[l+1,c_j(i)]}(t_{1/2}),h/2)$
\Else
\State compute QR fact. $\Lambda^{[l,i]}(t_{0})=Q^{[l,i]}(t_{0})R^{[l,i]}(t_{0})$
\State $\Lambda^{[l,i]}(t_0) \leftarrow Q^{[l,i]}(t_{0})$
\State $\Lambda^{[l+1,c_j(i)]}(t_0) \leftarrow Q^{[l+1,c_j(i)]}(t_{0})R^{[l,i]}(t_{0})$

\EndIf
\end{algorithmic} \end{algorithm}
\begin{algorithm}[H]
\caption{Propagation of top-node's tensor}
\label{alg:core}
\begin{algorithmic}[1]
\Require $\Psi[0,1](t_0)$
\Ensure $\Psi[0,1](t_1)$
\State $\Lambda^{[0,1]}(t_{1}) \leftarrow \mathrm{propagate}(\Lambda^{[0,1]}(t_{0}),h)$
\end{algorithmic} \end{algorithm}
\end{figure}

\subsection{Remarks}

The algorithm introduced above is a generalization of a previously
published projector-splitting integrator for TTNS with a single-layer
\citep{Lubich2014a,doi:10.1063/1.4982065}. Ref.~\citealp{ceruti2020time}
describes an algorithm for a general TTNS, which is identical to the
above algorithm with a single (either forward or backward) walk per
time-step. The main differences between the algorithm of Ref.\ \citealp{bauernfeind2019time}
and the one presented here are in the definition of the walk on the
tree and in the absence of a top-node, including it's separate propagation
routine.

While the TDVP applied to MPS has been demonstrated to be capable
of simulating dynamics in two-dimensional systems \citep{paeckel2019time},
a detailed analysis and comparison with other tensor network structures
is absent in the literature. In particular, the numerical stability
of the TDVP cannot be taken for granted \citep{yang2020time}, especially
when interactions between sites are long-ranged and not smoothly decaying,
as discussed in the following.

The application of TDVP formally requires the TTNS corresponding to
the initial condition to possess a full tree rank of $\mathbf{r}$.
However, many physical initial conditions of interest can be represented
with a low rank TTNS or even as a product state. If the initial condition
is not contained in the manifold of TTNS with tree rank of $\mathbf{r}$
due to rank deficiency, the TDVP doesn't provide a prescription for
how to choose and evolve the redundant parameters, which will gain
weight in the wavefunction representation at later times. Stability
and exactness of the dynamics under such circumstances is then dependent
on details of the implementation and the model. For the projector
splitting integrator, the initial rank-deficiency translates into
non-uniqueness of the matrix decompositions employed in the change
of isometrization. While the algorithm is not guaranteed to be exact
in this case, numerical experiments and prior applications of the
algorithm in one-dimensional systems indicate that it is generally
reliable even for product state initial conditions. As a check, one
may choose to regularize the initial condition by the addition of
weak noise, and test for invariance of the resulting dynamics at short
times. The initial evolution of redundant variational parameters depends
on arbitrary choices such as their initialization, the choice of regularization
(if applied), as well as the details of the linear algebra routines
used. Thus, different initializations of the same physical state may
not converge to the same solution \citep{Hinz_2016,manthe2015multi}.
Several approaches have been developed to address this problem. In
one-dimensional systems with nearest-neighbour interactions, the commonly
used two-site version of the TDVP algorithm of Ref. \citep{Haegeman2016}
is free of this issue, although this comes at the cost of breaking
unitarity of the evolution when the results cease to be close to the
exact solution. For generic interactions and arbitrary TTNS, a scheme
to optimally initialize redundant parameters was introduced \citep{manthe2015multi}.
However, this scheme requires the evaluation of an effective Hamiltonian
matrices for $\hat{H}^{2}$ and its compatibility with the integration
scheme employed here is an open question. Recently, another approach
based on a global basis expansion for MPS has been presented, and
should also be applicable to general TTNS \citep{yang2020time}.

Practically, we observe that the dependence of our results on non-optimal
initializations of redundant parameters systematically decreases with
increasing bond-dimension, which is also expected from the derivation
of the optimization scheme mentioned above. The dependence on initialization
becomes noticeable only when the wavefunction markedly departs from
the exact result, which provides an additional handle to access the
convergence of the method.

\section{\label{sec:Results}Results}

We first benchmark the method developed in this work by comparison
with exact results obtained for non-interacting fermions on a 2D lattice.
In the second stage we propagate a 2D system of hard-core bosons with
nearest neighbor interactions and compare our results to propagation
using MPS \citep{Zaletel2015}. The mapping of physical sites to the
respective tensor network structure is illustrated in Figs. \ref{fig:binary}
and \ref{fig:quaternary}. All calculations employ a regularization
of the initial product state, which consists of addition of white
noise sampled uniformly from the interval $[-10^{-20},10^{-20}]$
and subsequent renormalization of the TTNS.

\subsection{Non-interacting fermions}

\begin{figure}
\includegraphics{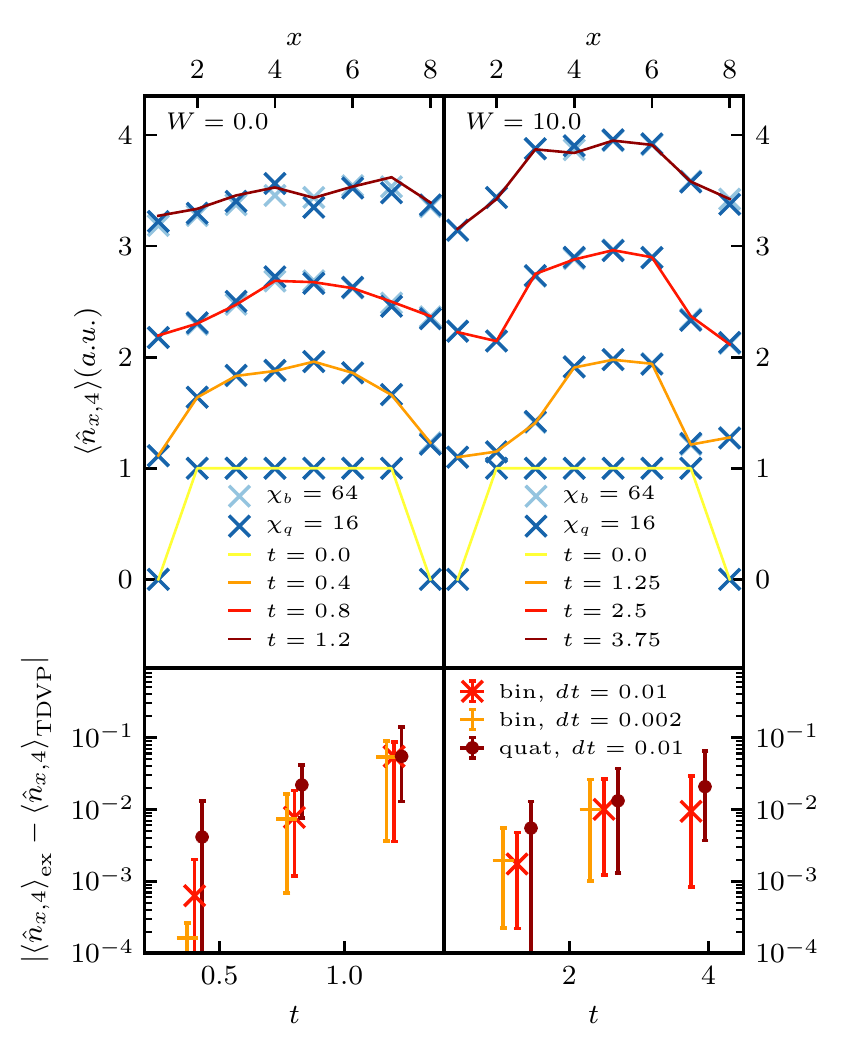}

\caption{\label{fig:NIF-Profiles}Density profiles of a central, horizontal
cut in the fourth row for a random product state configuration of
non-interacting fermions on a 8x8 lattice. \emph{Upper panels}: Profiles
for $W=0$ (left) and $W=10$ (right). Later times are spaced upwards
by 1 for readability. TDVP results for binary tensor network with
$\chi_{b}=64$ (light blue crosses) and quaternary tensor network
with $\chi_{b}=16$ (dark blue crosses), both with $dt=0.01$, shown
on top of exact results (solid lines). \emph{Lower panels}: The caps
of the error bars represent the maximal and minimal deviation of the
profiles in the above panels from the exact result for different bond-dimensions,
tensor network structures and time-steps.}
\end{figure}

\begin{figure}
\includegraphics{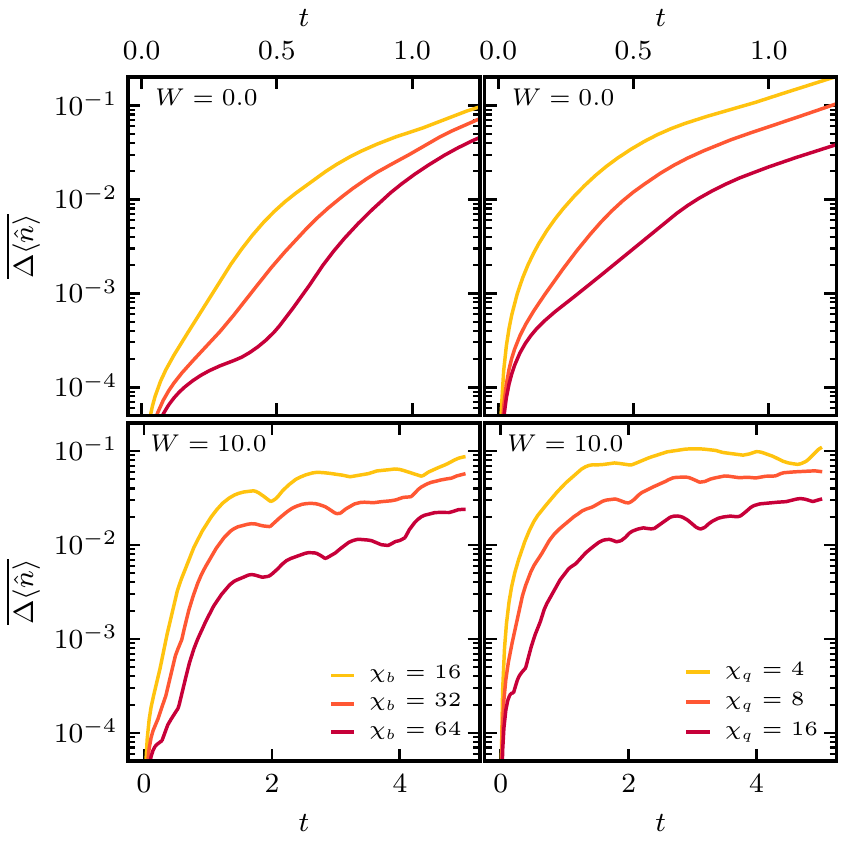}

\caption{\label{fig:NIF_results}Average deviation from exact $\left\langle \hat{n}\left(t\right)\right\rangle $
expectation value per site for non-interacting fermions on a clean
(top panels) and a disordered (bottom panels, $W=10$) 8x8 lattice
with open boundary conditions. Left panels are binary TTNS and right
panels are quaternary TTNS.The time step used is $dt=0.01$.}
\end{figure}

We compute the dynamics of non-interacting fermions on a 2D lattice
with on-site disorder
\begin{equation}
\hat{H}=J\sum_{<\boldsymbol{i},\boldsymbol{j}>}\left(\hat{c}_{\boldsymbol{i}}^{\dagger}\hat{c}_{\boldsymbol{j}}+\hat{c}_{\boldsymbol{j}}^{\dagger}\hat{c}_{\boldsymbol{i}}\right)+\sum_{\boldsymbol{i}}h_{\boldsymbol{i}}\left(\hat{c}_{\boldsymbol{i}}^{\dagger}\hat{c}_{\boldsymbol{i}}-\frac{1}{2}\right),\label{eq:H_NIF_F}
\end{equation}
where the index $\boldsymbol{i}=\left(x,y\right)$ indicates the position
of the fermion on the lattice, $\left\langle .\right\rangle $ is
a sum over nearest-neighbours, $h_{i}$ is drawn from a uniform distribution
$[-W,W]$ and $J=1$. All simulations use an identical initial state
which is a random product state at half-filling, and use a time step
$dt=0.01$. The tensor network state calculations employ the Jordan-Wigner
transformation of (\ref{eq:H_NIF_F})
\begin{multline}
\hat{H}=\sum_{<j,k>,j<k}\hat{S}_{j}^{+}\left(\prod_{j\leq l<k}2\hat{S}_{l}^{z}\right)\hat{S}_{k}^{-}\\
+\hat{S}_{j}^{-}\left(\prod_{j<l\leq k}2\hat{S}_{l}^{z}\right)\hat{S}_{k}^{+}+W\sum_{i}h_{i}\hat{S}_{i}^{z}.\label{eq:H_NIF_S}
\end{multline}
Different paths along which the sites are enumerated can be chosen,
and this choice potentially influences the performance of the TNS
algorithm. Here, we choose the path such that the Jordan-Wigner strings
span a minimal distance on the graph of the tree tensor network structure.
While solving the non-interacting problem in the fermionic representation
is trivial, the presence of Jordan-Wigner strings renders its solution
with tensor network states just as difficult as that of an interacting
problem. We compute the dynamics of this mode both for a clean system
($W=0$) and for one realization of a moderately strong quenched disorder
($W=10$). Two-dimensional non-interacting fermions show Anderson
localization at any finite disorder strength. While the localization
length may exceed the lattice dimensions chosen, disorder nonetheless
slows the growth of entanglement and should allow access to longer
timescales. Indeed, we observe good agreement for the density profiles,
$\hat{n}_{x,y}=\hat{c}_{x,y}^{\dagger}\hat{c}_{x,y}$ with $x,y\in\left[1,L\right]$,
along a horizontal cut of the lattice between the exact result and
data from both binary and quaternary TNS only up to times $t\leq1$
for the clean system, while longer times are accessible in the disordered
case (see Fig.\ \ref{fig:NIF-Profiles}). If the time-step is chosen
sufficiently small, errors associated to the linearization of Eq.~(\ref{eq:TDPSI_PT})
are negligible compared to inaccuracies related to the finite bond
dimensions at all but the earliest times (see lower panel of Fig.\ \ref{fig:NIF-Profiles}).
To get a more complete picture of the growth of errors with time as
well as their dependence on TNS structure and bond dimension, in Fig.~\ref{fig:NIF_results}
we show the average error in the expectation value of the local density
as a function of time. Both TNS structures show systematic improvement
with increasing bond dimension, and the error grows more mildly at
intermediate times in the disordered case. In both cases, smaller
deviations from exact results are achievable for binary TTNS than
for quaternary TTNS at the employed bond dimensions. We find that
a convergence criterion of an average error in the local density of
about $2$\,\% agrees well with the qualitative analysis of Fig.\ \ref{fig:NIF-Profiles}
and gives a good estimate of the times up to which the TNS results
are reliable.


\subsection{Hard-Core Bosons (XXZ model in 2D)}

\begin{figure}
\includegraphics[width=1\columnwidth]{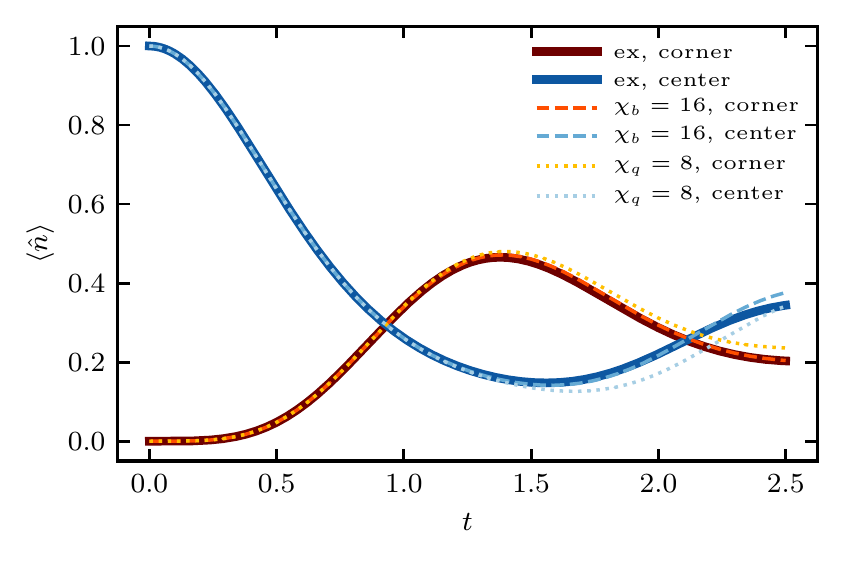}

\caption{\label{fig:HCB_benchmarks}Bosonic site density as a function of time
for a 4x4 lattice with the central 2x2 sites filled at $t=0$. Two
special sites are shown (corner and central). Exact results (solid
lines) and TNS results for binary (dashed lines) and quaternary (dotted
lines) TNS. Time step used for both panels is $dt=0.01$.}
\end{figure}

We consider the dynamics of hard-core bosons on a 2D lattice ,
\[
\hat{H}=-J\sum_{<\boldsymbol{i},\boldsymbol{j}>}\left(\hat{b}_{\boldsymbol{i}}^{\dagger}\hat{b}_{\boldsymbol{j}}+\hat{b}_{\boldsymbol{j}}^{\dagger}\hat{b}_{\boldsymbol{i}}\right)+V\sum_{<\boldsymbol{i},\boldsymbol{j}>}\hat{b}_{\boldsymbol{i}}^{\dagger}\hat{b}_{\boldsymbol{i}}\hat{b}_{\boldsymbol{j}}^{\dagger}\hat{b}_{\boldsymbol{j}}
\]
with nearest-neighbor interactions and we set $V=J=1$. We choose
an initial condition with a central square sublattice occupied and
all other lattice sites empty. This system and initial condition have
been studied previously in Ref.~\citealp{Zaletel2015} using MPS,
where results up to $tJ=2.0$ were presented for a square lattice
of linear length $L=14$. To establish the numerical exactness of
the algorithm for this non-integrable model, we compare the results
for the local bosonic density $\hat{n}_{x,y}=\hat{b}_{x,y}^{\dagger}\hat{b}_{x,y}$
with $x,y\in\left[1,L\right]$, for both binary and quaternary tensor
networks with exact diagonalization for a square lattice of linear
length $L=4$, with the central 4 lattice sites occupied (see Fig.~\ref{fig:HCB_benchmarks}).
Deviations from the exact result become noticeable only for times
$t\approx2$.

\begin{figure}
\includegraphics[width=1\columnwidth]{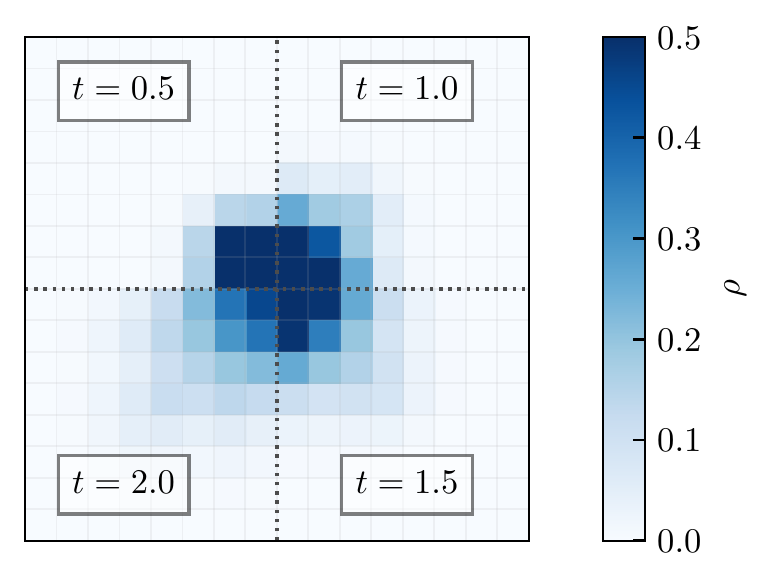}

\caption{\label{fig:Spreading} Spreading of hard-core boson density $\left\langle \hat{n}_{x,y}\right\rangle $,
initially occupying the central 4-by-4 sublattice of a square lattice
with $L=16$. Time step used is $dt=0.01$, and the scale is restricted
to a maximum of $n_{i}=0.5$ for clarity.}
\end{figure}

\begin{figure}
\includegraphics{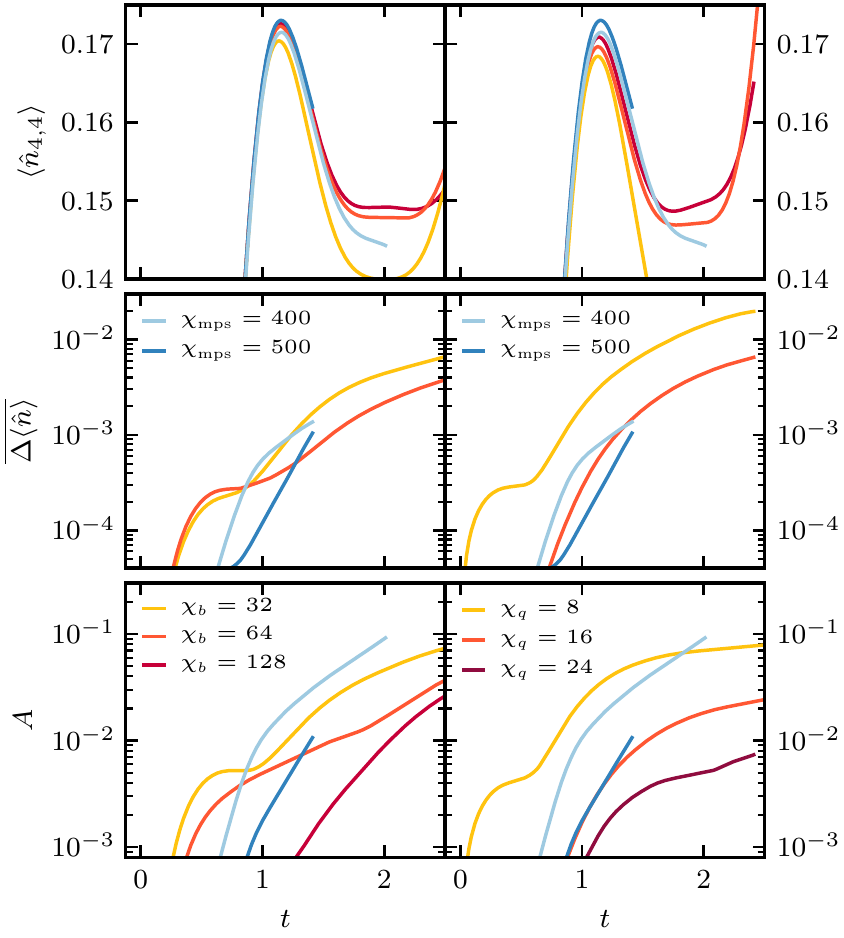}

\caption{\label{fig:Anisotropy}Measures of convergence for hard-core bosons
in 16x16 lattice for binary (left panels) and quaternary (right panels)
TTNS as well as MPS \citep{Zaletel2015} (all panels, blue shades).
\emph{Upper panels}: Bosonic density for the 4th left and 4th topmost
site. \emph{Middle panels}: Average deviation of the local bosonic
density with respect to best available result within the respective
TNS structure, for binary TTNS and MPS (left panel) as well as quaternary
TTNS and MPS (right panel). For $\chi_{\mathrm{mps}}=500$, the deviation
is reported with respect to $\chi_{b}=128$. . \emph{Bottom panels}:
Anisotropy (see text) of bosonic density. The time step used is $dt=0.01$.}
\end{figure}

Having established the validity of the algorithm, we investigate the
dynamics of an initial product state of a filled, central 4x4 sublattice
in a square lattice of a linear length $L=16$, see Fig.~\ref{fig:Spreading}.
In the upper panel of Fig.~\ref{fig:Anisotropy}, we focus on the
bosonic density for the site in the fourth row and fourth column of
the lattice. In contrast to the non-interacting model and the small
two-dimensional lattice discussed above, no exact results are available
for this interacting system and $L=16$. Therefore, the convergence
of the results is assessed by comparing the deviation of the local
density between different bond dimensions. All examined bond dimensions
agree well up to times $t\sim1.0$. For later times we see agreement
for all but the lowest bond dimensions in both quaternary and binary
TNS. However, \emph{quantitative} agreement (within a deviation of
0.001) up to $t=1.5$ only holds between the binary TNS results with
$\chi_{b}=128$, $\chi_{b}=64$ and the MPS results of Ref.~\citealp{Zaletel2015}
at $\chi=400$ and $\chi=500$. Since the accuracy of n-ary TTNS can
show site-dependence \citep{PhysRevB.87.125139}, we also report the
average density deviation with respect to the best available calculation
in the respective TNS structures in Fig.~\ref{fig:Anisotropy}. The
averaged density supports the observations made for a diagonal site
both quantitatively and qualitatively. Particularly, an average deviation
of $0.001$ is reached at $t=1.5$ for binary TNS, while quaternary
TNS saturate the threshold at $t=1.2$. The MPS results of Ref.~\citealp{Zaletel2015}
are converged to within this accuracy up to $t=1.3$, while the deviation
between the reference results of both binary TNS and MPS reach the
threshold at $t=1.4$.

Furthermore, since the Hamiltonian and the initial condition are isotropic,
distance from the exact solution can also be assessed by the anisotropy
$A(t)=$$\frac{1}{\sum_{x,y=1}^{L}n_{x,y}(t)}\sum_{x,y=1}^{L}\left|\hat{n}_{x,y}(t)-\hat{n}_{y,x}(t)\right|$
of the bosonic density, also reported in Fig.~\ref{fig:Anisotropy}.
We note however, that while the isotropy of the numerical solution
is required, it is not a sufficient condition for the solution to
be numerically exact. For both quaternary and binary TTNS, small anisotropies
($<0.3\%$) are obtained up to their respective convergence times.
In Ref.~\citealp{Zaletel2015}, an anisotropy of $4$\% was reported
at $t=2.0$ using MPS, a threshold which neither binary nor quaternary
TTNS saturate at the longest simulated times. Generally, the quaternary
TTNS has less anisotropic error since the partitioning of the lattice
through the tree structure is isotropic, although the result is less
tightly converged than the binary TTNS. Thus, anisotropy is only a
useful indicator of convergence when comparing TTNS of the same structure.
Given the small deviations in both anisotropy and local densities,
we consider our results to be numerically exact up $t=1.5$ for binary
TTNS with $\chi_{b}=128$ , and up to $t=1.2$ for quaternary TTNS
with $\chi_{q}=16$. The performance of the TDVP applied to binary
TTNS is thus comparable with the results of Ref.~\citealp{Zaletel2015},
providing the gain of better isotropy of the solution. Note that the
bond dimension used for MPS calculations do not correspond to the
current state-of-the-art, and larger bond dimension may be feasible
for binary TNS when using symmetries of the Hamiltonian. Due to the
lack of an exact solution to compare to, the convergence criterion
employed is significantly tighter than in the case of free fermions
to ensure quantitatively accurate results.


\section{\label{sec:Conclusion}Discussion}

In this work we have assessed the performance of TTNS for simulating
the dynamics of two-dimensional many-body lattice systems. We introduced
an algorithm based on the time-dependent variational principle for
arbitrary TTNS and benchmarked it on systems of non-interacting fermions
and interacting hard-core bosons in two dimensions, comparing the
performance to previously published results using matrix product states.
During the preparation of the manuscript we became aware of a recent
complementary work introducing a similar versions of the algorithm,
which were applied in rather different settings (as an impurity solver
\citep{bauernfeind2019time}, and in a more formal derivation of the
algorithm \citep{ceruti2020time}).

Currently, no efficient technique exists for exactly simulating the
non-equilibrium dynamics of interacting, two-dimensional quantum systems.
Despite recent progress, the timescales accessible by tensor network
techniques for such systems are generally extremely short. We have
found tree tensor networks to perform at least as well as matrix product
state techniques, with binary TTNS generally providing a more robust
performance than their quaternary counterparts. The issue of analyzing
the convergence, and thus ensuring the numerical exactness of the
computed result, was discussed. We believe the availability of an
alternative to matrix product states in the form of more general TTNS
is important and can offer additional insight in situations when slow
convergence is observed.

Our analysis has been mostly qualitative and a promising future avenue
is the exploration of the entanglement structure of out-of-equilibrium
states in 2D lattices . This will aid in the identification of optimal
tensor network structures in order to best exploit the increased flexibility
of TTNS, which already has proven to be important in applications
for zero-dimensional systems, such as impurity models and also for
molecular quantum dynamics \citep{wang2008coherent,wilner2013bistability,binder2018conformational,schroder2019tensor,bauernfeind2019time}.
The dynamics of one-dimensional systems quenched to a critical point
is another application where such an increased flexibility may be
of advantage. For critical systems in equilibrium, the multi-scale
entanglement renormalization (MERA) \citep{PhysRevLett.99.220405,aguado2008entanglement}
ansatz provides an efficient tensor network structure, which bears
resemblance with the $n$-ary tree structures employed here. However,
since a time-evolution approach for MERA is missing, it is interesting
to compare the performance of MPS and $n$-ary TTNS for critical systems
out-of-equilibrium. We leave such an investigation to a future work.
\begin{acknowledgments}
We thank Frank Pollmann for providing raw data from Ref.~\citep{Zaletel2015}.
BK and DRR acknowledge funding through the National Science Foundation
Grant No. CHE-1954791. YBL acknowledges support by the Israel Science
Foundation (grants No. 527/19 and 218/19).
\end{acknowledgments}

\bibliographystyle{aipnum4-1}
\bibliography{library_link,local}

\begin{thebibliography}{75}%
\makeatletter
\providecommand \@ifxundefined [1]{%
 \@ifx{#1\undefined}
}%
\providecommand \@ifnum [1]{%
 \ifnum #1\expandafter \@firstoftwo
 \else \expandafter \@secondoftwo
 \fi
}%
\providecommand \@ifx [1]{%
 \ifx #1\expandafter \@firstoftwo
 \else \expandafter \@secondoftwo
 \fi
}%
\providecommand \natexlab [1]{#1}%
\providecommand \enquote  [1]{``#1''}%
\providecommand \bibnamefont  [1]{#1}%
\providecommand \bibfnamefont [1]{#1}%
\providecommand \citenamefont [1]{#1}%
\providecommand \href@noop [0]{\@secondoftwo}%
\providecommand \href [0]{\begingroup \@sanitize@url \@href}%
\providecommand \@href[1]{\@@startlink{#1}\@@href}%
\providecommand \@@href[1]{\endgroup#1\@@endlink}%
\providecommand \@sanitize@url [0]{\catcode `\\12\catcode `\$12\catcode
  `\&12\catcode `\#12\catcode `\^12\catcode `\_12\catcode `\%12\relax}%
\providecommand \@@startlink[1]{}%
\providecommand \@@endlink[0]{}%
\providecommand \url  [0]{\begingroup\@sanitize@url \@url }%
\providecommand \@url [1]{\endgroup\@href {#1}{\urlprefix }}%
\providecommand \urlprefix  [0]{URL }%
\providecommand \Eprint [0]{\href }%
\providecommand \doibase [0]{http://dx.doi.org/}%
\providecommand \selectlanguage [0]{\@gobble}%
\providecommand \bibinfo  [0]{\@secondoftwo}%
\providecommand \bibfield  [0]{\@secondoftwo}%
\providecommand \translation [1]{[#1]}%
\providecommand \BibitemOpen [0]{}%
\providecommand \bibitemStop [0]{}%
\providecommand \bibitemNoStop [0]{.\EOS\space}%
\providecommand \EOS [0]{\spacefactor3000\relax}%
\providecommand \BibitemShut  [1]{\csname bibitem#1\endcsname}%
\let\auto@bib@innerbib\@empty
\bibitem [{\citenamefont {Calabrese}\ and\ \citenamefont
  {Cardy}(2005)}]{Calabrese_2005}%
  \BibitemOpen
  \bibfield  {author} {\bibinfo {author} {\bibfnamefont {P.}~\bibnamefont
  {Calabrese}}\ and\ \bibinfo {author} {\bibfnamefont {J.}~\bibnamefont
  {Cardy}},\ }\href {\doibase 10.1088/1742-5468/2005/04/p04010} {\bibfield
  {journal} {\bibinfo  {journal} {Journal of Statistical Mechanics: Theory and
  Experiment}\ }\textbf {\bibinfo {volume} {2005}},\ \bibinfo {pages} {P04010}
  (\bibinfo {year} {2005})}\BibitemShut {NoStop}%
\bibitem [{\citenamefont {De~Chiara}\ \emph {et~al.}(2006)\citenamefont
  {De~Chiara}, \citenamefont {Montangero}, \citenamefont {Calabrese},\ and\
  \citenamefont {Fazio}}]{chiara2006entanglement}%
  \BibitemOpen
  \bibfield  {author} {\bibinfo {author} {\bibfnamefont {G.}~\bibnamefont
  {De~Chiara}}, \bibinfo {author} {\bibfnamefont {S.}~\bibnamefont
  {Montangero}}, \bibinfo {author} {\bibfnamefont {P.}~\bibnamefont
  {Calabrese}}, \ and\ \bibinfo {author} {\bibfnamefont {R.}~\bibnamefont
  {Fazio}},\ }\href@noop {} {\bibfield  {journal} {\bibinfo  {journal} {Journal
  of Statistical Mechanics: Theory and Experiment}\ }\textbf {\bibinfo {volume}
  {2006}},\ \bibinfo {pages} {P03001} (\bibinfo {year} {2006})}\BibitemShut
  {NoStop}%
\bibitem [{\citenamefont {Alba}\ and\ \citenamefont
  {Calabrese}(2017)}]{Alba7947}%
  \BibitemOpen
  \bibfield  {author} {\bibinfo {author} {\bibfnamefont {V.}~\bibnamefont
  {Alba}}\ and\ \bibinfo {author} {\bibfnamefont {P.}~\bibnamefont
  {Calabrese}},\ }\href {\doibase 10.1073/pnas.1703516114} {\bibfield
  {journal} {\bibinfo  {journal} {Proceedings of the National Academy of
  Sciences}\ }\textbf {\bibinfo {volume} {114}},\ \bibinfo {pages} {7947}
  (\bibinfo {year} {2017})},\ \Eprint
  {http://arxiv.org/abs/https://www.pnas.org/content/114/30/7947.full.pdf}
  {https://www.pnas.org/content/114/30/7947.full.pdf} \BibitemShut {NoStop}%
\bibitem [{\citenamefont {Bloch}, \citenamefont {Dalibard},\ and\ \citenamefont
  {Zwerger}(2008)}]{RevModPhys.80.885}%
  \BibitemOpen
  \bibfield  {author} {\bibinfo {author} {\bibfnamefont {I.}~\bibnamefont
  {Bloch}}, \bibinfo {author} {\bibfnamefont {J.}~\bibnamefont {Dalibard}}, \
  and\ \bibinfo {author} {\bibfnamefont {W.}~\bibnamefont {Zwerger}},\ }\href
  {\doibase 10.1103/RevModPhys.80.885} {\bibfield  {journal} {\bibinfo
  {journal} {Rev. Mod. Phys.}\ }\textbf {\bibinfo {volume} {80}},\ \bibinfo
  {pages} {885} (\bibinfo {year} {2008})}\BibitemShut {NoStop}%
\bibitem [{\citenamefont {White}(1992)}]{PhysRevLett.69.2863}%
  \BibitemOpen
  \bibfield  {author} {\bibinfo {author} {\bibfnamefont {S.~R.}\ \bibnamefont
  {White}},\ }\href {\doibase 10.1103/PhysRevLett.69.2863} {\bibfield
  {journal} {\bibinfo  {journal} {Phys. Rev. Lett.}\ }\textbf {\bibinfo
  {volume} {69}},\ \bibinfo {pages} {2863} (\bibinfo {year}
  {1992})}\BibitemShut {NoStop}%
\bibitem [{\citenamefont {Hastings}(2007)}]{Hastings_2007}%
  \BibitemOpen
  \bibfield  {author} {\bibinfo {author} {\bibfnamefont {M.~B.}\ \bibnamefont
  {Hastings}},\ }\href {\doibase 10.1088/1742-5468/2007/08/p08024} {\bibfield
  {journal} {\bibinfo  {journal} {Journal of Statistical Mechanics: Theory and
  Experiment}\ }\textbf {\bibinfo {volume} {2007}},\ \bibinfo {pages} {P08024}
  (\bibinfo {year} {2007})}\BibitemShut {NoStop}%
\bibitem [{\citenamefont {Schollw{\"o}ck}(2011)}]{SCHOLLWOCK201196}%
  \BibitemOpen
  \bibfield  {author} {\bibinfo {author} {\bibfnamefont {U.}~\bibnamefont
  {Schollw{\"o}ck}},\ }\href {\doibase
  https://doi.org/10.1016/j.aop.2010.09.012} {\bibfield  {journal} {\bibinfo
  {journal} {Annals of Physics}\ }\textbf {\bibinfo {volume} {326}},\ \bibinfo
  {pages} {96 } (\bibinfo {year} {2011})},\ \bibinfo {note} {january 2011
  Special Issue}\BibitemShut {NoStop}%
\bibitem [{\citenamefont {Verstraete}\ and\ \citenamefont
  {Cirac}(2004)}]{verstraete2004renormalization}%
  \BibitemOpen
  \bibfield  {author} {\bibinfo {author} {\bibfnamefont {F.}~\bibnamefont
  {Verstraete}}\ and\ \bibinfo {author} {\bibfnamefont {J.~I.}\ \bibnamefont
  {Cirac}},\ }\href@noop {} {\enquote {\bibinfo {title} {Renormalization
  algorithms for quantum-many body systems in two and higher dimensions},}\ }
  (\bibinfo {year} {2004}),\ \Eprint {http://arxiv.org/abs/cond-mat/0407066}
  {arXiv:cond-mat/0407066 [cond-mat.str-el]} \BibitemShut {NoStop}%
\bibitem [{\citenamefont {Verstraete}\ \emph {et~al.}(2006)\citenamefont
  {Verstraete}, \citenamefont {Wolf}, \citenamefont {Perez-Garcia},\ and\
  \citenamefont {Cirac}}]{verstraete2006criticality}%
  \BibitemOpen
  \bibfield  {author} {\bibinfo {author} {\bibfnamefont {F.}~\bibnamefont
  {Verstraete}}, \bibinfo {author} {\bibfnamefont {M.~M.}\ \bibnamefont
  {Wolf}}, \bibinfo {author} {\bibfnamefont {D.}~\bibnamefont {Perez-Garcia}},
  \ and\ \bibinfo {author} {\bibfnamefont {J.~I.}\ \bibnamefont {Cirac}},\
  }\href@noop {} {\bibfield  {journal} {\bibinfo  {journal} {Physical review
  letters}\ }\textbf {\bibinfo {volume} {96}},\ \bibinfo {pages} {220601}
  (\bibinfo {year} {2006})}\BibitemShut {NoStop}%
\bibitem [{\citenamefont {Lubasch}, \citenamefont {Cirac},\ and\ \citenamefont
  {Banuls}(2014)}]{lubasch2014unifying}%
  \BibitemOpen
  \bibfield  {author} {\bibinfo {author} {\bibfnamefont {M.}~\bibnamefont
  {Lubasch}}, \bibinfo {author} {\bibfnamefont {J.~I.}\ \bibnamefont {Cirac}},
  \ and\ \bibinfo {author} {\bibfnamefont {M.-C.}\ \bibnamefont {Banuls}},\
  }\href@noop {} {\bibfield  {journal} {\bibinfo  {journal} {New Journal of
  Physics}\ }\textbf {\bibinfo {volume} {16}},\ \bibinfo {pages} {033014}
  (\bibinfo {year} {2014})}\BibitemShut {NoStop}%
\bibitem [{\citenamefont {Ran}\ \emph {et~al.}(2017)\citenamefont {Ran},
  \citenamefont {Tirrito}, \citenamefont {Peng}, \citenamefont {Chen},
  \citenamefont {Tagliacozzo}, \citenamefont {Su},\ and\ \citenamefont
  {Lewenstein}}]{ran2017lecture}%
  \BibitemOpen
  \bibfield  {author} {\bibinfo {author} {\bibfnamefont {S.-J.}\ \bibnamefont
  {Ran}}, \bibinfo {author} {\bibfnamefont {E.}~\bibnamefont {Tirrito}},
  \bibinfo {author} {\bibfnamefont {C.}~\bibnamefont {Peng}}, \bibinfo {author}
  {\bibfnamefont {X.}~\bibnamefont {Chen}}, \bibinfo {author} {\bibfnamefont
  {L.}~\bibnamefont {Tagliacozzo}}, \bibinfo {author} {\bibfnamefont
  {G.}~\bibnamefont {Su}}, \ and\ \bibinfo {author} {\bibfnamefont
  {M.}~\bibnamefont {Lewenstein}},\ }\href@noop {} {\bibfield  {journal}
  {\bibinfo  {journal} {arXiv preprint arXiv:1708.09213}\ } (\bibinfo {year}
  {2017})}\BibitemShut {NoStop}%
\bibitem [{\citenamefont {Zheng}\ \emph {et~al.}(2017)\citenamefont {Zheng},
  \citenamefont {Chung}, \citenamefont {Corboz}, \citenamefont {Ehlers},
  \citenamefont {Qin}, \citenamefont {Noack}, \citenamefont {Shi},
  \citenamefont {White}, \citenamefont {Zhang},\ and\ \citenamefont
  {Chan}}]{zheng2017stripe}%
  \BibitemOpen
  \bibfield  {author} {\bibinfo {author} {\bibfnamefont {B.-X.}\ \bibnamefont
  {Zheng}}, \bibinfo {author} {\bibfnamefont {C.-M.}\ \bibnamefont {Chung}},
  \bibinfo {author} {\bibfnamefont {P.}~\bibnamefont {Corboz}}, \bibinfo
  {author} {\bibfnamefont {G.}~\bibnamefont {Ehlers}}, \bibinfo {author}
  {\bibfnamefont {M.-P.}\ \bibnamefont {Qin}}, \bibinfo {author} {\bibfnamefont
  {R.~M.}\ \bibnamefont {Noack}}, \bibinfo {author} {\bibfnamefont
  {H.}~\bibnamefont {Shi}}, \bibinfo {author} {\bibfnamefont {S.~R.}\
  \bibnamefont {White}}, \bibinfo {author} {\bibfnamefont {S.}~\bibnamefont
  {Zhang}}, \ and\ \bibinfo {author} {\bibfnamefont {G.~K.-L.}\ \bibnamefont
  {Chan}},\ }\href@noop {} {\bibfield  {journal} {\bibinfo  {journal}
  {Science}\ }\textbf {\bibinfo {volume} {358}},\ \bibinfo {pages} {1155}
  (\bibinfo {year} {2017})}\BibitemShut {NoStop}%
\bibitem [{\citenamefont {Pi{\v{z}}orn}, \citenamefont {Wang},\ and\
  \citenamefont {Verstraete}(2011)}]{pivzorn2011time}%
  \BibitemOpen
  \bibfield  {author} {\bibinfo {author} {\bibfnamefont {I.}~\bibnamefont
  {Pi{\v{z}}orn}}, \bibinfo {author} {\bibfnamefont {L.}~\bibnamefont {Wang}},
  \ and\ \bibinfo {author} {\bibfnamefont {F.}~\bibnamefont {Verstraete}},\
  }\href@noop {} {\bibfield  {journal} {\bibinfo  {journal} {Physical Review
  A}\ }\textbf {\bibinfo {volume} {83}},\ \bibinfo {pages} {052321} (\bibinfo
  {year} {2011})}\BibitemShut {NoStop}%
\bibitem [{\citenamefont {Kshetrimayum}, \citenamefont {Goihl},\ and\
  \citenamefont {Eisert}(2019)}]{kshetrimayum2019time}%
  \BibitemOpen
  \bibfield  {author} {\bibinfo {author} {\bibfnamefont {A.}~\bibnamefont
  {Kshetrimayum}}, \bibinfo {author} {\bibfnamefont {M.}~\bibnamefont {Goihl}},
  \ and\ \bibinfo {author} {\bibfnamefont {J.}~\bibnamefont {Eisert}},\
  }\href@noop {} {\bibfield  {journal} {\bibinfo  {journal} {arXiv preprint
  arXiv:1910.11359}\ } (\bibinfo {year} {2019})}\BibitemShut {NoStop}%
\bibitem [{\citenamefont {Hubig}\ \emph {et~al.}(2019)\citenamefont {Hubig},
  \citenamefont {Bohrdt}, \citenamefont {Knap}, \citenamefont {Grusdt},\ and\
  \citenamefont {Cirac}}]{hubig2019evaluation}%
  \BibitemOpen
  \bibfield  {author} {\bibinfo {author} {\bibfnamefont {C.}~\bibnamefont
  {Hubig}}, \bibinfo {author} {\bibfnamefont {A.}~\bibnamefont {Bohrdt}},
  \bibinfo {author} {\bibfnamefont {M.}~\bibnamefont {Knap}}, \bibinfo {author}
  {\bibfnamefont {F.}~\bibnamefont {Grusdt}}, \ and\ \bibinfo {author}
  {\bibfnamefont {J.~I.}\ \bibnamefont {Cirac}},\ }\href@noop {} {\bibfield
  {journal} {\bibinfo  {journal} {arXiv preprint arXiv:1911.01159}\ } (\bibinfo
  {year} {2019})}\BibitemShut {NoStop}%
\bibitem [{\citenamefont {Zaletel}\ and\ \citenamefont
  {Pollmann}(2019)}]{zaletel2019isometric}%
  \BibitemOpen
  \bibfield  {author} {\bibinfo {author} {\bibfnamefont {M.~P.}\ \bibnamefont
  {Zaletel}}\ and\ \bibinfo {author} {\bibfnamefont {F.}~\bibnamefont
  {Pollmann}},\ }\href@noop {} {\bibfield  {journal} {\bibinfo  {journal}
  {arXiv preprint arXiv:1902.05100}\ } (\bibinfo {year} {2019})}\BibitemShut
  {NoStop}%
\bibitem [{\citenamefont {Crosswhite}, \citenamefont {Doherty},\ and\
  \citenamefont {Vidal}(2008)}]{crosswhite2008applying}%
  \BibitemOpen
  \bibfield  {author} {\bibinfo {author} {\bibfnamefont {G.~M.}\ \bibnamefont
  {Crosswhite}}, \bibinfo {author} {\bibfnamefont {A.~C.}\ \bibnamefont
  {Doherty}}, \ and\ \bibinfo {author} {\bibfnamefont {G.}~\bibnamefont
  {Vidal}},\ }\href@noop {} {\bibfield  {journal} {\bibinfo  {journal}
  {Physical Review B}\ }\textbf {\bibinfo {volume} {78}},\ \bibinfo {pages}
  {035116} (\bibinfo {year} {2008})}\BibitemShut {NoStop}%
\bibitem [{\citenamefont {Fr{\"o}wis}, \citenamefont {Nebendahl},\ and\
  \citenamefont {D{\"u}r}(2010)}]{frowis2010tensor}%
  \BibitemOpen
  \bibfield  {author} {\bibinfo {author} {\bibfnamefont {F.}~\bibnamefont
  {Fr{\"o}wis}}, \bibinfo {author} {\bibfnamefont {V.}~\bibnamefont
  {Nebendahl}}, \ and\ \bibinfo {author} {\bibfnamefont {W.}~\bibnamefont
  {D{\"u}r}},\ }\href@noop {} {\bibfield  {journal} {\bibinfo  {journal}
  {Physical Review A}\ }\textbf {\bibinfo {volume} {81}},\ \bibinfo {pages}
  {062337} (\bibinfo {year} {2010})}\BibitemShut {NoStop}%
\bibitem [{\citenamefont {Stoudenmire}\ and\ \citenamefont
  {White}(2012)}]{stoudenmire2012studying}%
  \BibitemOpen
  \bibfield  {author} {\bibinfo {author} {\bibfnamefont {E.~M.}\ \bibnamefont
  {Stoudenmire}}\ and\ \bibinfo {author} {\bibfnamefont {S.~R.}\ \bibnamefont
  {White}},\ }\href@noop {} {\bibfield  {journal} {\bibinfo  {journal} {Annu.
  Rev. Condens. Matter Phys.}\ }\textbf {\bibinfo {volume} {3}},\ \bibinfo
  {pages} {111} (\bibinfo {year} {2012})}\BibitemShut {NoStop}%
\bibitem [{\citenamefont {Zaletel}\ \emph {et~al.}(2015)\citenamefont
  {Zaletel}, \citenamefont {Mong}, \citenamefont {Karrasch}, \citenamefont
  {Moore},\ and\ \citenamefont {Pollmann}}]{Zaletel2015}%
  \BibitemOpen
  \bibfield  {author} {\bibinfo {author} {\bibfnamefont {M.~P.}\ \bibnamefont
  {Zaletel}}, \bibinfo {author} {\bibfnamefont {R.~S.~K.}\ \bibnamefont
  {Mong}}, \bibinfo {author} {\bibfnamefont {C.}~\bibnamefont {Karrasch}},
  \bibinfo {author} {\bibfnamefont {J.~E.}\ \bibnamefont {Moore}}, \ and\
  \bibinfo {author} {\bibfnamefont {F.}~\bibnamefont {Pollmann}},\ }\href
  {\doibase 10.1103/PhysRevB.91.165112} {\bibfield  {journal} {\bibinfo
  {journal} {Physical Review B - Condensed Matter and Materials Physics}\
  }\textbf {\bibinfo {volume} {91}},\ \bibinfo {pages} {1} (\bibinfo {year}
  {2015})},\ \Eprint {http://arxiv.org/abs/1407.1832} {arXiv:1407.1832}
  \BibitemShut {NoStop}%
\bibitem [{\citenamefont {Paeckel}\ \emph {et~al.}(2019)\citenamefont
  {Paeckel}, \citenamefont {K{\"o}hler}, \citenamefont {Swoboda}, \citenamefont
  {Manmana}, \citenamefont {Schollw{\"o}ck},\ and\ \citenamefont
  {Hubig}}]{paeckel2019time}%
  \BibitemOpen
  \bibfield  {author} {\bibinfo {author} {\bibfnamefont {S.}~\bibnamefont
  {Paeckel}}, \bibinfo {author} {\bibfnamefont {T.}~\bibnamefont {K{\"o}hler}},
  \bibinfo {author} {\bibfnamefont {A.}~\bibnamefont {Swoboda}}, \bibinfo
  {author} {\bibfnamefont {S.~R.}\ \bibnamefont {Manmana}}, \bibinfo {author}
  {\bibfnamefont {U.}~\bibnamefont {Schollw{\"o}ck}}, \ and\ \bibinfo {author}
  {\bibfnamefont {C.}~\bibnamefont {Hubig}},\ }\href@noop {} {\bibfield
  {journal} {\bibinfo  {journal} {Annals of Physics}\ }\textbf {\bibinfo
  {volume} {411}},\ \bibinfo {pages} {167998} (\bibinfo {year}
  {2019})}\BibitemShut {NoStop}%
\bibitem [{\citenamefont {Doggen}\ \emph {et~al.}(2020)\citenamefont {Doggen},
  \citenamefont {Gornyi}, \citenamefont {Mirlin},\ and\ \citenamefont
  {Polyakov}}]{doggen2020slow}%
  \BibitemOpen
  \bibfield  {author} {\bibinfo {author} {\bibfnamefont {E.~V.~H.}\
  \bibnamefont {Doggen}}, \bibinfo {author} {\bibfnamefont {I.~V.}\
  \bibnamefont {Gornyi}}, \bibinfo {author} {\bibfnamefont {A.~D.}\
  \bibnamefont {Mirlin}}, \ and\ \bibinfo {author} {\bibfnamefont {D.~G.}\
  \bibnamefont {Polyakov}},\ }\href@noop {} {\enquote {\bibinfo {title} {Slow
  many-body delocalization beyond one dimension},}\ } (\bibinfo {year}
  {2020}),\ \Eprint {http://arxiv.org/abs/2002.07635} {arXiv:2002.07635
  [cond-mat.dis-nn]} \BibitemShut {NoStop}%
\bibitem [{\citenamefont {Carleo}\ and\ \citenamefont
  {Troyer}(2017)}]{carleo2017solving}%
  \BibitemOpen
  \bibfield  {author} {\bibinfo {author} {\bibfnamefont {G.}~\bibnamefont
  {Carleo}}\ and\ \bibinfo {author} {\bibfnamefont {M.}~\bibnamefont
  {Troyer}},\ }\href@noop {} {\bibfield  {journal} {\bibinfo  {journal}
  {Science}\ }\textbf {\bibinfo {volume} {355}},\ \bibinfo {pages} {602}
  (\bibinfo {year} {2017})}\BibitemShut {NoStop}%
\bibitem [{\citenamefont {Schmitt}\ and\ \citenamefont
  {Heyl}(2019)}]{schmitt2019quantum}%
  \BibitemOpen
  \bibfield  {author} {\bibinfo {author} {\bibfnamefont {M.}~\bibnamefont
  {Schmitt}}\ and\ \bibinfo {author} {\bibfnamefont {M.}~\bibnamefont {Heyl}},\
  }\href@noop {} {\enquote {\bibinfo {title} {Quantum many-body dynamics in two
  dimensions with artificial neural networks},}\ } (\bibinfo {year} {2019}),\
  \Eprint {http://arxiv.org/abs/1912.08828} {arXiv:1912.08828
  [cond-mat.str-el]} \BibitemShut {NoStop}%
\bibitem [{\citenamefont {L\'{o}pez-Guti\'{e}rrez}\ and\ \citenamefont
  {Mendl}(2019)}]{lpezgutirrez2019real}%
  \BibitemOpen
  \bibfield  {author} {\bibinfo {author} {\bibfnamefont {I.}~\bibnamefont
  {L\'{o}pez-Guti\'{e}rrez}}\ and\ \bibinfo {author} {\bibfnamefont {C.~B.}\
  \bibnamefont {Mendl}},\ }\href@noop {} {\enquote {\bibinfo {title} {Real time
  evolution with neural-network quantum states},}\ } (\bibinfo {year} {2019}),\
  \Eprint {http://arxiv.org/abs/1912.08831} {arXiv:1912.08831
  [cond-mat.dis-nn]} \BibitemShut {NoStop}%
\bibitem [{\citenamefont {Rizzi}\ \emph {et~al.}(2010)\citenamefont {Rizzi},
  \citenamefont {Montangero}, \citenamefont {Silvi}, \citenamefont
  {Giovannetti},\ and\ \citenamefont {Fazio}}]{Rizzi_2010}%
  \BibitemOpen
  \bibfield  {author} {\bibinfo {author} {\bibfnamefont {M.}~\bibnamefont
  {Rizzi}}, \bibinfo {author} {\bibfnamefont {S.}~\bibnamefont {Montangero}},
  \bibinfo {author} {\bibfnamefont {P.}~\bibnamefont {Silvi}}, \bibinfo
  {author} {\bibfnamefont {V.}~\bibnamefont {Giovannetti}}, \ and\ \bibinfo
  {author} {\bibfnamefont {R.}~\bibnamefont {Fazio}},\ }\href {\doibase
  10.1088/1367-2630/12/7/075018} {\bibfield  {journal} {\bibinfo  {journal}
  {New Journal of Physics}\ }\textbf {\bibinfo {volume} {12}},\ \bibinfo
  {pages} {075018} (\bibinfo {year} {2010})}\BibitemShut {NoStop}%
\bibitem [{\citenamefont {Tsai}, \citenamefont {Chen},\ and\ \citenamefont
  {Lin}(2020)}]{tsai2020tensor}%
  \BibitemOpen
  \bibfield  {author} {\bibinfo {author} {\bibfnamefont {Z.-L.}\ \bibnamefont
  {Tsai}}, \bibinfo {author} {\bibfnamefont {P.}~\bibnamefont {Chen}}, \ and\
  \bibinfo {author} {\bibfnamefont {Y.-C.}\ \bibnamefont {Lin}},\ }\href@noop
  {} {\bibfield  {journal} {\bibinfo  {journal} {The European Physical Journal
  B}\ }\textbf {\bibinfo {volume} {93}},\ \bibinfo {pages} {1} (\bibinfo {year}
  {2020})}\BibitemShut {NoStop}%
\bibitem [{\citenamefont {Shi}, \citenamefont {Duan},\ and\ \citenamefont
  {Vidal}(2006)}]{PhysRevA.74.022320}%
  \BibitemOpen
  \bibfield  {author} {\bibinfo {author} {\bibfnamefont {Y.-Y.}\ \bibnamefont
  {Shi}}, \bibinfo {author} {\bibfnamefont {L.-M.}\ \bibnamefont {Duan}}, \
  and\ \bibinfo {author} {\bibfnamefont {G.}~\bibnamefont {Vidal}},\ }\href
  {\doibase 10.1103/PhysRevA.74.022320} {\bibfield  {journal} {\bibinfo
  {journal} {Phys. Rev. A}\ }\textbf {\bibinfo {volume} {74}},\ \bibinfo
  {pages} {022320} (\bibinfo {year} {2006})}\BibitemShut {NoStop}%
\bibitem [{\citenamefont {Tagliacozzo}, \citenamefont {Evenbly},\ and\
  \citenamefont {Vidal}(2009)}]{Tagliacozzo2009}%
  \BibitemOpen
  \bibfield  {author} {\bibinfo {author} {\bibfnamefont {L.}~\bibnamefont
  {Tagliacozzo}}, \bibinfo {author} {\bibfnamefont {G.}~\bibnamefont
  {Evenbly}}, \ and\ \bibinfo {author} {\bibfnamefont {G.}~\bibnamefont
  {Vidal}},\ }\href {\doibase 10.1103/PhysRevB.80.235127} {\bibfield  {journal}
  {\bibinfo  {journal} {Physical Review B - Condensed Matter and Materials
  Physics}\ }\textbf {\bibinfo {volume} {80}},\ \bibinfo {pages} {1} (\bibinfo
  {year} {2009})},\ \Eprint {http://arxiv.org/abs/0903.5017} {arXiv:0903.5017}
  \BibitemShut {NoStop}%
\bibitem [{\citenamefont {Murg}\ \emph {et~al.}(2010)\citenamefont {Murg},
  \citenamefont {Verstraete}, \citenamefont {Legeza},\ and\ \citenamefont
  {Noack}}]{PhysRevB.82.205105}%
  \BibitemOpen
  \bibfield  {author} {\bibinfo {author} {\bibfnamefont {V.}~\bibnamefont
  {Murg}}, \bibinfo {author} {\bibfnamefont {F.}~\bibnamefont {Verstraete}},
  \bibinfo {author} {\bibfnamefont {O.}~\bibnamefont {Legeza}}, \ and\ \bibinfo
  {author} {\bibfnamefont {R.~M.}\ \bibnamefont {Noack}},\ }\href {\doibase
  10.1103/PhysRevB.82.205105} {\bibfield  {journal} {\bibinfo  {journal} {Phys.
  Rev. B}\ }\textbf {\bibinfo {volume} {82}},\ \bibinfo {pages} {205105}
  (\bibinfo {year} {2010})}\BibitemShut {NoStop}%
\bibitem [{\citenamefont {Li}, \citenamefont {von Delft},\ and\ \citenamefont
  {Xiang}(2012)}]{PhysRevB.86.195137}%
  \BibitemOpen
  \bibfield  {author} {\bibinfo {author} {\bibfnamefont {W.}~\bibnamefont
  {Li}}, \bibinfo {author} {\bibfnamefont {J.}~\bibnamefont {von Delft}}, \
  and\ \bibinfo {author} {\bibfnamefont {T.}~\bibnamefont {Xiang}},\ }\href
  {\doibase 10.1103/PhysRevB.86.195137} {\bibfield  {journal} {\bibinfo
  {journal} {Phys. Rev. B}\ }\textbf {\bibinfo {volume} {86}},\ \bibinfo
  {pages} {195137} (\bibinfo {year} {2012})}\BibitemShut {NoStop}%
\bibitem [{\citenamefont {Gerster}\ \emph {et~al.}(2014)\citenamefont
  {Gerster}, \citenamefont {Silvi}, \citenamefont {Rizzi}, \citenamefont
  {Fazio}, \citenamefont {Calarco},\ and\ \citenamefont
  {Montangero}}]{Gerster2014}%
  \BibitemOpen
  \bibfield  {author} {\bibinfo {author} {\bibfnamefont {M.}~\bibnamefont
  {Gerster}}, \bibinfo {author} {\bibfnamefont {P.}~\bibnamefont {Silvi}},
  \bibinfo {author} {\bibfnamefont {M.}~\bibnamefont {Rizzi}}, \bibinfo
  {author} {\bibfnamefont {R.}~\bibnamefont {Fazio}}, \bibinfo {author}
  {\bibfnamefont {T.}~\bibnamefont {Calarco}}, \ and\ \bibinfo {author}
  {\bibfnamefont {S.}~\bibnamefont {Montangero}},\ }\href {\doibase
  10.1103/PhysRevB.90.125154} {\bibfield  {journal} {\bibinfo  {journal}
  {Physical Review B - Condensed Matter and Materials Physics}\ }\textbf
  {\bibinfo {volume} {90}},\ \bibinfo {pages} {1} (\bibinfo {year} {2014})},\
  \Eprint {http://arxiv.org/abs/1406.2666} {arXiv:1406.2666} \BibitemShut
  {NoStop}%
\bibitem [{\citenamefont {Gerster}\ \emph {et~al.}(2017)\citenamefont
  {Gerster}, \citenamefont {Rizzi}, \citenamefont {Silvi}, \citenamefont
  {Dalmonte},\ and\ \citenamefont {Montangero}}]{PhysRevB.96.195123}%
  \BibitemOpen
  \bibfield  {author} {\bibinfo {author} {\bibfnamefont {M.}~\bibnamefont
  {Gerster}}, \bibinfo {author} {\bibfnamefont {M.}~\bibnamefont {Rizzi}},
  \bibinfo {author} {\bibfnamefont {P.}~\bibnamefont {Silvi}}, \bibinfo
  {author} {\bibfnamefont {M.}~\bibnamefont {Dalmonte}}, \ and\ \bibinfo
  {author} {\bibfnamefont {S.}~\bibnamefont {Montangero}},\ }\href {\doibase
  10.1103/PhysRevB.96.195123} {\bibfield  {journal} {\bibinfo  {journal} {Phys.
  Rev. B}\ }\textbf {\bibinfo {volume} {96}},\ \bibinfo {pages} {195123}
  (\bibinfo {year} {2017})}\BibitemShut {NoStop}%
\bibitem [{\citenamefont {Chepiga}\ and\ \citenamefont
  {White}(2019)}]{chepiga2019comb}%
  \BibitemOpen
  \bibfield  {author} {\bibinfo {author} {\bibfnamefont {N.}~\bibnamefont
  {Chepiga}}\ and\ \bibinfo {author} {\bibfnamefont {S.~R.}\ \bibnamefont
  {White}},\ }\href@noop {} {\bibfield  {journal} {\bibinfo  {journal}
  {Physical Review B}\ }\textbf {\bibinfo {volume} {99}},\ \bibinfo {pages}
  {235426} (\bibinfo {year} {2019})}\BibitemShut {NoStop}%
\bibitem [{\citenamefont {Milsted}\ \emph {et~al.}(2019)\citenamefont
  {Milsted}, \citenamefont {Ganahl}, \citenamefont {Leichenauer}, \citenamefont
  {Hidary},\ and\ \citenamefont {Vidal}}]{milsted2019tensornetwork}%
  \BibitemOpen
  \bibfield  {author} {\bibinfo {author} {\bibfnamefont {A.}~\bibnamefont
  {Milsted}}, \bibinfo {author} {\bibfnamefont {M.}~\bibnamefont {Ganahl}},
  \bibinfo {author} {\bibfnamefont {S.}~\bibnamefont {Leichenauer}}, \bibinfo
  {author} {\bibfnamefont {J.}~\bibnamefont {Hidary}}, \ and\ \bibinfo {author}
  {\bibfnamefont {G.}~\bibnamefont {Vidal}},\ }\href@noop {} {\enquote
  {\bibinfo {title} {Tensornetwork on tensorflow: A spin chain application
  using tree tensor networks},}\ } (\bibinfo {year} {2019}),\ \Eprint
  {http://arxiv.org/abs/1905.01331} {arXiv:1905.01331 [cond-mat.str-el]}
  \BibitemShut {NoStop}%
\bibitem [{\citenamefont {Macaluso}\ \emph {et~al.}(2020)\citenamefont
  {Macaluso}, \citenamefont {Comparin}, \citenamefont {Umucal\ifmmode \imath
  \else~\i \fi{}lar}, \citenamefont {Gerster}, \citenamefont {Montangero},
  \citenamefont {Rizzi},\ and\ \citenamefont
  {Carusotto}}]{PhysRevResearch.2.013145}%
  \BibitemOpen
  \bibfield  {author} {\bibinfo {author} {\bibfnamefont {E.}~\bibnamefont
  {Macaluso}}, \bibinfo {author} {\bibfnamefont {T.}~\bibnamefont {Comparin}},
  \bibinfo {author} {\bibfnamefont {R.~O.}\ \bibnamefont {Umucal\ifmmode \imath
  \else~\i \fi{}lar}}, \bibinfo {author} {\bibfnamefont {M.}~\bibnamefont
  {Gerster}}, \bibinfo {author} {\bibfnamefont {S.}~\bibnamefont {Montangero}},
  \bibinfo {author} {\bibfnamefont {M.}~\bibnamefont {Rizzi}}, \ and\ \bibinfo
  {author} {\bibfnamefont {I.}~\bibnamefont {Carusotto}},\ }\href {\doibase
  10.1103/PhysRevResearch.2.013145} {\bibfield  {journal} {\bibinfo  {journal}
  {Phys. Rev. Research}\ }\textbf {\bibinfo {volume} {2}},\ \bibinfo {pages}
  {013145} (\bibinfo {year} {2020})}\BibitemShut {NoStop}%
\bibitem [{\citenamefont {Nakatani}\ and\ \citenamefont
  {Chan}(2013)}]{Nakatani2013}%
  \BibitemOpen
  \bibfield  {author} {\bibinfo {author} {\bibfnamefont {N.}~\bibnamefont
  {Nakatani}}\ and\ \bibinfo {author} {\bibfnamefont {G.~K.~L.}\ \bibnamefont
  {Chan}},\ }\href {\doibase 10.1063/1.4798639} {\bibfield  {journal} {\bibinfo
   {journal} {Journal of Chemical Physics}\ }\textbf {\bibinfo {volume} {138}}
  (\bibinfo {year} {2013}),\ 10.1063/1.4798639},\ \Eprint
  {http://arxiv.org/abs/1302.2298} {arXiv:1302.2298} \BibitemShut {NoStop}%
\bibitem [{\citenamefont {Murg}\ \emph {et~al.}(2015)\citenamefont {Murg},
  \citenamefont {Verstraete}, \citenamefont {Schneider}, \citenamefont {Nagy},\
  and\ \citenamefont {Legeza}}]{Murg2015}%
  \BibitemOpen
  \bibfield  {author} {\bibinfo {author} {\bibfnamefont {V.}~\bibnamefont
  {Murg}}, \bibinfo {author} {\bibfnamefont {F.}~\bibnamefont {Verstraete}},
  \bibinfo {author} {\bibfnamefont {R.}~\bibnamefont {Schneider}}, \bibinfo
  {author} {\bibfnamefont {P.~R.}\ \bibnamefont {Nagy}}, \ and\ \bibinfo
  {author} {\bibnamefont {Legeza}},\ }\href {\doibase 10.1021/ct501187j}
  {\bibfield  {journal} {\bibinfo  {journal} {Journal of Chemical Theory and
  Computation}\ }\textbf {\bibinfo {volume} {11}},\ \bibinfo {pages} {1027}
  (\bibinfo {year} {2015})}\BibitemShut {NoStop}%
\bibitem [{\citenamefont {Gunst}\ \emph {et~al.}(2018)\citenamefont {Gunst},
  \citenamefont {Verstraete}, \citenamefont {Wouters}, \citenamefont {Legeza},\
  and\ \citenamefont {Van~Neck}}]{gunst2018t3ns}%
  \BibitemOpen
  \bibfield  {author} {\bibinfo {author} {\bibfnamefont {K.}~\bibnamefont
  {Gunst}}, \bibinfo {author} {\bibfnamefont {F.}~\bibnamefont {Verstraete}},
  \bibinfo {author} {\bibfnamefont {S.}~\bibnamefont {Wouters}}, \bibinfo
  {author} {\bibfnamefont {O.}~\bibnamefont {Legeza}}, \ and\ \bibinfo {author}
  {\bibfnamefont {D.}~\bibnamefont {Van~Neck}},\ }\href@noop {} {\bibfield
  {journal} {\bibinfo  {journal} {Journal of chemical theory and computation}\
  }\textbf {\bibinfo {volume} {14}},\ \bibinfo {pages} {2026} (\bibinfo {year}
  {2018})}\BibitemShut {NoStop}%
\bibitem [{\citenamefont {Meyer}, \citenamefont {Manthe},\ and\ \citenamefont
  {Cederbaum}(1990)}]{meyer1990multi}%
  \BibitemOpen
  \bibfield  {author} {\bibinfo {author} {\bibfnamefont {H.-D.}\ \bibnamefont
  {Meyer}}, \bibinfo {author} {\bibfnamefont {U.}~\bibnamefont {Manthe}}, \
  and\ \bibinfo {author} {\bibfnamefont {L.~S.}\ \bibnamefont {Cederbaum}},\
  }\href@noop {} {\bibfield  {journal} {\bibinfo  {journal} {Chemical Physics
  Letters}\ }\textbf {\bibinfo {volume} {165}},\ \bibinfo {pages} {73}
  (\bibinfo {year} {1990})}\BibitemShut {NoStop}%
\bibitem [{\citenamefont {Beck}\ \emph {et~al.}(2000)\citenamefont {Beck},
  \citenamefont {J{\"a}ckle}, \citenamefont {Worth},\ and\ \citenamefont
  {Meyer}}]{beck2000multiconfiguration}%
  \BibitemOpen
  \bibfield  {author} {\bibinfo {author} {\bibfnamefont {M.~H.}\ \bibnamefont
  {Beck}}, \bibinfo {author} {\bibfnamefont {A.}~\bibnamefont {J{\"a}ckle}},
  \bibinfo {author} {\bibfnamefont {G.~A.}\ \bibnamefont {Worth}}, \ and\
  \bibinfo {author} {\bibfnamefont {H.-D.}\ \bibnamefont {Meyer}},\ }\href@noop
  {} {\bibfield  {journal} {\bibinfo  {journal} {Physics reports}\ }\textbf
  {\bibinfo {volume} {324}},\ \bibinfo {pages} {1} (\bibinfo {year}
  {2000})}\BibitemShut {NoStop}%
\bibitem [{\citenamefont {Wang}\ and\ \citenamefont
  {Thoss}(2003)}]{wang2003multilayer}%
  \BibitemOpen
  \bibfield  {author} {\bibinfo {author} {\bibfnamefont {H.}~\bibnamefont
  {Wang}}\ and\ \bibinfo {author} {\bibfnamefont {M.}~\bibnamefont {Thoss}},\
  }\href@noop {} {\bibfield  {journal} {\bibinfo  {journal} {The Journal of
  chemical physics}\ }\textbf {\bibinfo {volume} {119}},\ \bibinfo {pages}
  {1289} (\bibinfo {year} {2003})}\BibitemShut {NoStop}%
\bibitem [{\citenamefont {Dirac}(1930)}]{dirac_1930}%
  \BibitemOpen
  \bibfield  {author} {\bibinfo {author} {\bibfnamefont {P.~A.~M.}\
  \bibnamefont {Dirac}},\ }\href {\doibase 10.1017/S0305004100016108}
  {\bibfield  {journal} {\bibinfo  {journal} {Mathematical Proceedings of the
  Cambridge Philosophical Society}\ }\textbf {\bibinfo {volume} {26}},\
  \bibinfo {pages} {376} (\bibinfo {year} {1930})}\BibitemShut {NoStop}%
\bibitem [{\citenamefont {Frenkel}\ \emph {et~al.}(1934)\citenamefont {Frenkel}
  \emph {et~al.}}]{frenkel1934wave}%
  \BibitemOpen
  \bibfield  {author} {\bibinfo {author} {\bibfnamefont {I.}~\bibnamefont
  {Frenkel}} \emph {et~al.},\ }\href@noop {} {\  (\bibinfo {year}
  {1934})}\BibitemShut {NoStop}%
\bibitem [{\citenamefont {Haegeman}\ \emph {et~al.}(2011)\citenamefont
  {Haegeman}, \citenamefont {Cirac}, \citenamefont {Osborne}, \citenamefont
  {Pi\ifmmode~\check{z}\else \v{z}\fi{}orn}, \citenamefont {Verschelde},\ and\
  \citenamefont {Verstraete}}]{PhysRevLett.107.070601}%
  \BibitemOpen
  \bibfield  {author} {\bibinfo {author} {\bibfnamefont {J.}~\bibnamefont
  {Haegeman}}, \bibinfo {author} {\bibfnamefont {J.~I.}\ \bibnamefont {Cirac}},
  \bibinfo {author} {\bibfnamefont {T.~J.}\ \bibnamefont {Osborne}}, \bibinfo
  {author} {\bibfnamefont {I.}~\bibnamefont {Pi\ifmmode~\check{z}\else
  \v{z}\fi{}orn}}, \bibinfo {author} {\bibfnamefont {H.}~\bibnamefont
  {Verschelde}}, \ and\ \bibinfo {author} {\bibfnamefont {F.}~\bibnamefont
  {Verstraete}},\ }\href {\doibase 10.1103/PhysRevLett.107.070601} {\bibfield
  {journal} {\bibinfo  {journal} {Phys. Rev. Lett.}\ }\textbf {\bibinfo
  {volume} {107}},\ \bibinfo {pages} {070601} (\bibinfo {year}
  {2011})}\BibitemShut {NoStop}%
\bibitem [{\citenamefont {Koffel}, \citenamefont {Lewenstein},\ and\
  \citenamefont {Tagliacozzo}(2012)}]{PhysRevLett.109.267203}%
  \BibitemOpen
  \bibfield  {author} {\bibinfo {author} {\bibfnamefont {T.}~\bibnamefont
  {Koffel}}, \bibinfo {author} {\bibfnamefont {M.}~\bibnamefont {Lewenstein}},
  \ and\ \bibinfo {author} {\bibfnamefont {L.}~\bibnamefont {Tagliacozzo}},\
  }\href {\doibase 10.1103/PhysRevLett.109.267203} {\bibfield  {journal}
  {\bibinfo  {journal} {Phys. Rev. Lett.}\ }\textbf {\bibinfo {volume} {109}},\
  \bibinfo {pages} {267203} (\bibinfo {year} {2012})}\BibitemShut {NoStop}%
\bibitem [{\citenamefont {Hauke}\ and\ \citenamefont
  {Tagliacozzo}(2013)}]{PhysRevLett.111.207202}%
  \BibitemOpen
  \bibfield  {author} {\bibinfo {author} {\bibfnamefont {P.}~\bibnamefont
  {Hauke}}\ and\ \bibinfo {author} {\bibfnamefont {L.}~\bibnamefont
  {Tagliacozzo}},\ }\href {\doibase 10.1103/PhysRevLett.111.207202} {\bibfield
  {journal} {\bibinfo  {journal} {Phys. Rev. Lett.}\ }\textbf {\bibinfo
  {volume} {111}},\ \bibinfo {pages} {207202} (\bibinfo {year}
  {2013})}\BibitemShut {NoStop}%
\bibitem [{\citenamefont {Lubich}, \citenamefont {Oseledets},\ and\
  \citenamefont {Vandereycken}(2014)}]{Lubich2014}%
  \BibitemOpen
  \bibfield  {author} {\bibinfo {author} {\bibfnamefont {C.}~\bibnamefont
  {Lubich}}, \bibinfo {author} {\bibfnamefont {I.~V.}\ \bibnamefont
  {Oseledets}}, \ and\ \bibinfo {author} {\bibfnamefont {B.}~\bibnamefont
  {Vandereycken}},\ }\href {\doibase 10.1137/140976546} {\bibfield  {journal}
  {\bibinfo  {journal} {ArXiv e-prints}\ }\textbf {\bibinfo {volume} {53}},\
  \bibinfo {pages} {917} (\bibinfo {year} {2014})},\ \Eprint
  {http://arxiv.org/abs/1407.2042} {arXiv:1407.2042} \BibitemShut {NoStop}%
\bibitem [{\citenamefont {Haegeman}\ \emph {et~al.}(2016)\citenamefont
  {Haegeman}, \citenamefont {Lubich}, \citenamefont {Oseledets}, \citenamefont
  {Vandereycken},\ and\ \citenamefont {Verstraete}}]{Haegeman2016}%
  \BibitemOpen
  \bibfield  {author} {\bibinfo {author} {\bibfnamefont {J.}~\bibnamefont
  {Haegeman}}, \bibinfo {author} {\bibfnamefont {C.}~\bibnamefont {Lubich}},
  \bibinfo {author} {\bibfnamefont {I.}~\bibnamefont {Oseledets}}, \bibinfo
  {author} {\bibfnamefont {B.}~\bibnamefont {Vandereycken}}, \ and\ \bibinfo
  {author} {\bibfnamefont {F.}~\bibnamefont {Verstraete}},\ }\href {\doibase
  10.1103/PhysRevB.94.165116} {\bibfield  {journal} {\bibinfo  {journal}
  {Physical Review B}\ }\textbf {\bibinfo {volume} {94}},\ \bibinfo {pages} {1}
  (\bibinfo {year} {2016})},\ \Eprint {http://arxiv.org/abs/1408.5056}
  {arXiv:1408.5056} \BibitemShut {NoStop}%
\bibitem [{\citenamefont {Leviatan}\ \emph {et~al.}(2017)\citenamefont
  {Leviatan}, \citenamefont {Pollmann}, \citenamefont {Bardarson},
  \citenamefont {Huse},\ and\ \citenamefont {Altman}}]{leviatan2017quantum}%
  \BibitemOpen
  \bibfield  {author} {\bibinfo {author} {\bibfnamefont {E.}~\bibnamefont
  {Leviatan}}, \bibinfo {author} {\bibfnamefont {F.}~\bibnamefont {Pollmann}},
  \bibinfo {author} {\bibfnamefont {J.~H.}\ \bibnamefont {Bardarson}}, \bibinfo
  {author} {\bibfnamefont {D.~A.}\ \bibnamefont {Huse}}, \ and\ \bibinfo
  {author} {\bibfnamefont {E.}~\bibnamefont {Altman}},\ }\href@noop {}
  {\enquote {\bibinfo {title} {Quantum thermalization dynamics with
  matrix-product states},}\ } (\bibinfo {year} {2017}),\ \Eprint
  {http://arxiv.org/abs/1702.08894} {arXiv:1702.08894 [cond-mat.stat-mech]}
  \BibitemShut {NoStop}%
\bibitem [{\citenamefont {Kloss}, \citenamefont {Lev},\ and\ \citenamefont
  {Reichman}(2018)}]{PhysRevB.97.024307}%
  \BibitemOpen
  \bibfield  {author} {\bibinfo {author} {\bibfnamefont {B.}~\bibnamefont
  {Kloss}}, \bibinfo {author} {\bibfnamefont {Y.~B.}\ \bibnamefont {Lev}}, \
  and\ \bibinfo {author} {\bibfnamefont {D.}~\bibnamefont {Reichman}},\ }\href
  {\doibase 10.1103/PhysRevB.97.024307} {\bibfield  {journal} {\bibinfo
  {journal} {Phys. Rev. B}\ }\textbf {\bibinfo {volume} {97}},\ \bibinfo
  {pages} {024307} (\bibinfo {year} {2018})}\BibitemShut {NoStop}%
\bibitem [{\citenamefont {Krumnow}\ \emph {et~al.}(2016)\citenamefont
  {Krumnow}, \citenamefont {Veis}, \citenamefont {Legeza},\ and\ \citenamefont
  {Eisert}}]{PhysRevLett.117.210402}%
  \BibitemOpen
  \bibfield  {author} {\bibinfo {author} {\bibfnamefont {C.}~\bibnamefont
  {Krumnow}}, \bibinfo {author} {\bibfnamefont {L.}~\bibnamefont {Veis}},
  \bibinfo {author} {\bibfnamefont {O.}~\bibnamefont {Legeza}}, \ and\ \bibinfo
  {author} {\bibfnamefont {J.}~\bibnamefont {Eisert}},\ }\href {\doibase
  10.1103/PhysRevLett.117.210402} {\bibfield  {journal} {\bibinfo  {journal}
  {Phys. Rev. Lett.}\ }\textbf {\bibinfo {volume} {117}},\ \bibinfo {pages}
  {210402} (\bibinfo {year} {2016})}\BibitemShut {NoStop}%
\bibitem [{\citenamefont {Krumnow}, \citenamefont {Eisert},\ and\ \citenamefont
  {Legeza}(2019)}]{krumnow2019overcoming}%
  \BibitemOpen
  \bibfield  {author} {\bibinfo {author} {\bibfnamefont {C.}~\bibnamefont
  {Krumnow}}, \bibinfo {author} {\bibfnamefont {J.}~\bibnamefont {Eisert}}, \
  and\ \bibinfo {author} {\bibfnamefont {O.}~\bibnamefont {Legeza}},\
  }\href@noop {} {\enquote {\bibinfo {title} {Towards overcoming the
  entanglement barrier when simulating long-time evolution},}\ } (\bibinfo
  {year} {2019}),\ \Eprint {http://arxiv.org/abs/1904.11999} {arXiv:1904.11999
  [cond-mat.stat-mech]} \BibitemShut {NoStop}%
\bibitem [{\citenamefont {Rams}\ and\ \citenamefont
  {Zwolak}(2020)}]{PhysRevLett.124.137701}%
  \BibitemOpen
  \bibfield  {author} {\bibinfo {author} {\bibfnamefont {M.~M.}\ \bibnamefont
  {Rams}}\ and\ \bibinfo {author} {\bibfnamefont {M.}~\bibnamefont {Zwolak}},\
  }\href {\doibase 10.1103/PhysRevLett.124.137701} {\bibfield  {journal}
  {\bibinfo  {journal} {Phys. Rev. Lett.}\ }\textbf {\bibinfo {volume} {124}},\
  \bibinfo {pages} {137701} (\bibinfo {year} {2020})}\BibitemShut {NoStop}%
\bibitem [{\citenamefont {Hauschild}\ \emph {et~al.}(2018)\citenamefont
  {Hauschild}, \citenamefont {Leviatan}, \citenamefont {Bardarson},
  \citenamefont {Altman}, \citenamefont {Zaletel},\ and\ \citenamefont
  {Pollmann}}]{PhysRevB.98.235163}%
  \BibitemOpen
  \bibfield  {author} {\bibinfo {author} {\bibfnamefont {J.}~\bibnamefont
  {Hauschild}}, \bibinfo {author} {\bibfnamefont {E.}~\bibnamefont {Leviatan}},
  \bibinfo {author} {\bibfnamefont {J.~H.}\ \bibnamefont {Bardarson}}, \bibinfo
  {author} {\bibfnamefont {E.}~\bibnamefont {Altman}}, \bibinfo {author}
  {\bibfnamefont {M.~P.}\ \bibnamefont {Zaletel}}, \ and\ \bibinfo {author}
  {\bibfnamefont {F.}~\bibnamefont {Pollmann}},\ }\href {\doibase
  10.1103/PhysRevB.98.235163} {\bibfield  {journal} {\bibinfo  {journal} {Phys.
  Rev. B}\ }\textbf {\bibinfo {volume} {98}},\ \bibinfo {pages} {235163}
  (\bibinfo {year} {2018})}\BibitemShut {NoStop}%
\bibitem [{\citenamefont {Surace}, \citenamefont {Piani},\ and\ \citenamefont
  {Tagliacozzo}(2019)}]{PhysRevB.99.235115}%
  \BibitemOpen
  \bibfield  {author} {\bibinfo {author} {\bibfnamefont {J.}~\bibnamefont
  {Surace}}, \bibinfo {author} {\bibfnamefont {M.}~\bibnamefont {Piani}}, \
  and\ \bibinfo {author} {\bibfnamefont {L.}~\bibnamefont {Tagliacozzo}},\
  }\href {\doibase 10.1103/PhysRevB.99.235115} {\bibfield  {journal} {\bibinfo
  {journal} {Phys. Rev. B}\ }\textbf {\bibinfo {volume} {99}},\ \bibinfo
  {pages} {235115} (\bibinfo {year} {2019})}\BibitemShut {NoStop}%
\bibitem [{\citenamefont {White}\ \emph {et~al.}(2018)\citenamefont {White},
  \citenamefont {Zaletel}, \citenamefont {Mong},\ and\ \citenamefont
  {Refael}}]{PhysRevB.97.035127}%
  \BibitemOpen
  \bibfield  {author} {\bibinfo {author} {\bibfnamefont {C.~D.}\ \bibnamefont
  {White}}, \bibinfo {author} {\bibfnamefont {M.}~\bibnamefont {Zaletel}},
  \bibinfo {author} {\bibfnamefont {R.~S.~K.}\ \bibnamefont {Mong}}, \ and\
  \bibinfo {author} {\bibfnamefont {G.}~\bibnamefont {Refael}},\ }\href
  {\doibase 10.1103/PhysRevB.97.035127} {\bibfield  {journal} {\bibinfo
  {journal} {Phys. Rev. B}\ }\textbf {\bibinfo {volume} {97}},\ \bibinfo
  {pages} {035127} (\bibinfo {year} {2018})}\BibitemShut {NoStop}%
\bibitem [{\citenamefont {Rakovszky}, \citenamefont {von Keyserlingk},\ and\
  \citenamefont {Pollmann}(2020)}]{rakovszky2020dissipationassisted}%
  \BibitemOpen
  \bibfield  {author} {\bibinfo {author} {\bibfnamefont {T.}~\bibnamefont
  {Rakovszky}}, \bibinfo {author} {\bibfnamefont {C.~W.}\ \bibnamefont {von
  Keyserlingk}}, \ and\ \bibinfo {author} {\bibfnamefont {F.}~\bibnamefont
  {Pollmann}},\ }\href@noop {} {\enquote {\bibinfo {title}
  {Dissipation-assisted operator evolution method for capturing hydrodynamic
  transport},}\ } (\bibinfo {year} {2020}),\ \Eprint
  {http://arxiv.org/abs/2004.05177} {arXiv:2004.05177 [cond-mat.str-el]}
  \BibitemShut {NoStop}%
\bibitem [{\citenamefont {Lubich}(2014)}]{Lubich2014a}%
  \BibitemOpen
  \bibfield  {author} {\bibinfo {author} {\bibfnamefont {C.}~\bibnamefont
  {Lubich}},\ }\href {\doibase 10.1093/amrx/abv006} {\bibfield  {journal}
  {\bibinfo  {journal} {Applied Mathematics Research eXpress}\ }\textbf
  {\bibinfo {volume} {2015}},\ \bibinfo {pages} {311} (\bibinfo {year}
  {2014})}\BibitemShut {NoStop}%
\bibitem [{\citenamefont {Ferris}(2013)}]{PhysRevB.87.125139}%
  \BibitemOpen
  \bibfield  {author} {\bibinfo {author} {\bibfnamefont {A.~J.}\ \bibnamefont
  {Ferris}},\ }\href {\doibase 10.1103/PhysRevB.87.125139} {\bibfield
  {journal} {\bibinfo  {journal} {Phys. Rev. B}\ }\textbf {\bibinfo {volume}
  {87}},\ \bibinfo {pages} {125139} (\bibinfo {year} {2013})}\BibitemShut
  {NoStop}%
\bibitem [{\citenamefont {Schr{\"o}der}\ \emph {et~al.}(2019)\citenamefont
  {Schr{\"o}der}, \citenamefont {Turban}, \citenamefont {Musser}, \citenamefont
  {Hine},\ and\ \citenamefont {Chin}}]{schroder2019tensor}%
  \BibitemOpen
  \bibfield  {author} {\bibinfo {author} {\bibfnamefont {F.~A.}\ \bibnamefont
  {Schr{\"o}der}}, \bibinfo {author} {\bibfnamefont {D.~H.}\ \bibnamefont
  {Turban}}, \bibinfo {author} {\bibfnamefont {A.~J.}\ \bibnamefont {Musser}},
  \bibinfo {author} {\bibfnamefont {N.~D.}\ \bibnamefont {Hine}}, \ and\
  \bibinfo {author} {\bibfnamefont {A.~W.}\ \bibnamefont {Chin}},\ }\href@noop
  {} {\bibfield  {journal} {\bibinfo  {journal} {Nature communications}\
  }\textbf {\bibinfo {volume} {10}},\ \bibinfo {pages} {1} (\bibinfo {year}
  {2019})}\BibitemShut {NoStop}%
\bibitem [{\citenamefont {Bauernfeind}\ and\ \citenamefont
  {Aichhorn}(2020)}]{bauernfeind2019time}%
  \BibitemOpen
  \bibfield  {author} {\bibinfo {author} {\bibfnamefont {D.}~\bibnamefont
  {Bauernfeind}}\ and\ \bibinfo {author} {\bibfnamefont {M.}~\bibnamefont
  {Aichhorn}},\ }\href {\doibase 10.21468/SciPostPhys.8.2.024} {\bibfield
  {journal} {\bibinfo  {journal} {SciPost Phys.}\ }\textbf {\bibinfo {volume}
  {8}},\ \bibinfo {pages} {24} (\bibinfo {year} {2020})}\BibitemShut {NoStop}%
\bibitem [{\citenamefont {Ceruti}, \citenamefont {Lubich},\ and\ \citenamefont
  {Walach}(2020)}]{ceruti2020time}%
  \BibitemOpen
  \bibfield  {author} {\bibinfo {author} {\bibfnamefont {G.}~\bibnamefont
  {Ceruti}}, \bibinfo {author} {\bibfnamefont {C.}~\bibnamefont {Lubich}}, \
  and\ \bibinfo {author} {\bibfnamefont {H.}~\bibnamefont {Walach}},\
  }\href@noop {} {\bibfield  {journal} {\bibinfo  {journal} {arXiv preprint
  arXiv:2002.11392}\ } (\bibinfo {year} {2020})}\BibitemShut {NoStop}%
\bibitem [{\citenamefont {Nagaj}\ \emph {et~al.}(2008)\citenamefont {Nagaj},
  \citenamefont {Farhi}, \citenamefont {Goldstone}, \citenamefont {Shor},\ and\
  \citenamefont {Sylvester}}]{PhysRevB.77.214431}%
  \BibitemOpen
  \bibfield  {author} {\bibinfo {author} {\bibfnamefont {D.}~\bibnamefont
  {Nagaj}}, \bibinfo {author} {\bibfnamefont {E.}~\bibnamefont {Farhi}},
  \bibinfo {author} {\bibfnamefont {J.}~\bibnamefont {Goldstone}}, \bibinfo
  {author} {\bibfnamefont {P.}~\bibnamefont {Shor}}, \ and\ \bibinfo {author}
  {\bibfnamefont {I.}~\bibnamefont {Sylvester}},\ }\href {\doibase
  10.1103/PhysRevB.77.214431} {\bibfield  {journal} {\bibinfo  {journal} {Phys.
  Rev. B}\ }\textbf {\bibinfo {volume} {77}},\ \bibinfo {pages} {214431}
  (\bibinfo {year} {2008})}\BibitemShut {NoStop}%
\bibitem [{\citenamefont {Uschmajew}\ and\ \citenamefont
  {Vandereycken}(2013)}]{uschmajew2013geometry}%
  \BibitemOpen
  \bibfield  {author} {\bibinfo {author} {\bibfnamefont {A.}~\bibnamefont
  {Uschmajew}}\ and\ \bibinfo {author} {\bibfnamefont {B.}~\bibnamefont
  {Vandereycken}},\ }\href@noop {} {\bibfield  {journal} {\bibinfo  {journal}
  {Linear Algebra and its Applications}\ }\textbf {\bibinfo {volume} {439}},\
  \bibinfo {pages} {133} (\bibinfo {year} {2013})}\BibitemShut {NoStop}%
\bibitem [{\citenamefont {Lubich}\ \emph {et~al.}(2013)\citenamefont {Lubich},
  \citenamefont {Rohwedder}, \citenamefont {Schneider},\ and\ \citenamefont
  {Vandereycken}}]{lubich2013dynamical}%
  \BibitemOpen
  \bibfield  {author} {\bibinfo {author} {\bibfnamefont {C.}~\bibnamefont
  {Lubich}}, \bibinfo {author} {\bibfnamefont {T.}~\bibnamefont {Rohwedder}},
  \bibinfo {author} {\bibfnamefont {R.}~\bibnamefont {Schneider}}, \ and\
  \bibinfo {author} {\bibfnamefont {B.}~\bibnamefont {Vandereycken}},\
  }\href@noop {} {\bibfield  {journal} {\bibinfo  {journal} {SIAM Journal on
  Matrix Analysis and Applications}\ }\textbf {\bibinfo {volume} {34}},\
  \bibinfo {pages} {470} (\bibinfo {year} {2013})}\BibitemShut {NoStop}%
\bibitem [{\citenamefont {Kloss}, \citenamefont {Burghardt},\ and\
  \citenamefont {Lubich}(2017)}]{doi:10.1063/1.4982065}%
  \BibitemOpen
  \bibfield  {author} {\bibinfo {author} {\bibfnamefont {B.}~\bibnamefont
  {Kloss}}, \bibinfo {author} {\bibfnamefont {I.}~\bibnamefont {Burghardt}}, \
  and\ \bibinfo {author} {\bibfnamefont {C.}~\bibnamefont {Lubich}},\ }\href
  {\doibase 10.1063/1.4982065} {\bibfield  {journal} {\bibinfo  {journal} {The
  Journal of Chemical Physics}\ }\textbf {\bibinfo {volume} {146}},\ \bibinfo
  {pages} {174107} (\bibinfo {year} {2017})},\ \Eprint
  {http://arxiv.org/abs/https://doi.org/10.1063/1.4982065}
  {https://doi.org/10.1063/1.4982065} \BibitemShut {NoStop}%
\bibitem [{\citenamefont {Yang}\ and\ \citenamefont
  {White}(2020)}]{yang2020time}%
  \BibitemOpen
  \bibfield  {author} {\bibinfo {author} {\bibfnamefont {M.}~\bibnamefont
  {Yang}}\ and\ \bibinfo {author} {\bibfnamefont {S.~R.}\ \bibnamefont
  {White}},\ }\href@noop {} {\enquote {\bibinfo {title} {Time dependent
  variational principle with ancillary krylov subspace},}\ } (\bibinfo {year}
  {2020}),\ \Eprint {http://arxiv.org/abs/2005.06104} {arXiv:2005.06104
  [cond-mat.str-el]} \BibitemShut {NoStop}%
\bibitem [{\citenamefont {Hinz}, \citenamefont {Bauch},\ and\ \citenamefont
  {Bonitz}(2016)}]{Hinz_2016}%
  \BibitemOpen
  \bibfield  {author} {\bibinfo {author} {\bibfnamefont {C.~M.}\ \bibnamefont
  {Hinz}}, \bibinfo {author} {\bibfnamefont {S.}~\bibnamefont {Bauch}}, \ and\
  \bibinfo {author} {\bibfnamefont {M.}~\bibnamefont {Bonitz}},\ }\href
  {\doibase 10.1088/1742-6596/696/1/012009} {\bibfield  {journal} {\bibinfo
  {journal} {Journal of Physics: Conference Series}\ }\textbf {\bibinfo
  {volume} {696}},\ \bibinfo {pages} {012009} (\bibinfo {year}
  {2016})}\BibitemShut {NoStop}%
\bibitem [{\citenamefont {Manthe}(2015)}]{manthe2015multi}%
  \BibitemOpen
  \bibfield  {author} {\bibinfo {author} {\bibfnamefont {U.}~\bibnamefont
  {Manthe}},\ }\href@noop {} {\bibfield  {journal} {\bibinfo  {journal} {The
  Journal of chemical physics}\ }\textbf {\bibinfo {volume} {142}},\ \bibinfo
  {pages} {244109} (\bibinfo {year} {2015})}\BibitemShut {NoStop}%
\bibitem [{\citenamefont {Wang}\ and\ \citenamefont
  {Thoss}(2008)}]{wang2008coherent}%
  \BibitemOpen
  \bibfield  {author} {\bibinfo {author} {\bibfnamefont {H.}~\bibnamefont
  {Wang}}\ and\ \bibinfo {author} {\bibfnamefont {M.}~\bibnamefont {Thoss}},\
  }\href@noop {} {\bibfield  {journal} {\bibinfo  {journal} {New Journal of
  Physics}\ }\textbf {\bibinfo {volume} {10}},\ \bibinfo {pages} {115005}
  (\bibinfo {year} {2008})}\BibitemShut {NoStop}%
\bibitem [{\citenamefont {Wilner}\ \emph {et~al.}(2013)\citenamefont {Wilner},
  \citenamefont {Wang}, \citenamefont {Cohen}, \citenamefont {Thoss},\ and\
  \citenamefont {Rabani}}]{wilner2013bistability}%
  \BibitemOpen
  \bibfield  {author} {\bibinfo {author} {\bibfnamefont {E.~Y.}\ \bibnamefont
  {Wilner}}, \bibinfo {author} {\bibfnamefont {H.}~\bibnamefont {Wang}},
  \bibinfo {author} {\bibfnamefont {G.}~\bibnamefont {Cohen}}, \bibinfo
  {author} {\bibfnamefont {M.}~\bibnamefont {Thoss}}, \ and\ \bibinfo {author}
  {\bibfnamefont {E.}~\bibnamefont {Rabani}},\ }\href@noop {} {\bibfield
  {journal} {\bibinfo  {journal} {Physical Review B}\ }\textbf {\bibinfo
  {volume} {88}},\ \bibinfo {pages} {045137} (\bibinfo {year}
  {2013})}\BibitemShut {NoStop}%
\bibitem [{\citenamefont {Binder}, \citenamefont {Lauvergnat},\ and\
  \citenamefont {Burghardt}(2018)}]{binder2018conformational}%
  \BibitemOpen
  \bibfield  {author} {\bibinfo {author} {\bibfnamefont {R.}~\bibnamefont
  {Binder}}, \bibinfo {author} {\bibfnamefont {D.}~\bibnamefont {Lauvergnat}},
  \ and\ \bibinfo {author} {\bibfnamefont {I.}~\bibnamefont {Burghardt}},\
  }\href@noop {} {\bibfield  {journal} {\bibinfo  {journal} {Physical review
  letters}\ }\textbf {\bibinfo {volume} {120}},\ \bibinfo {pages} {227401}
  (\bibinfo {year} {2018})}\BibitemShut {NoStop}%
\bibitem [{\citenamefont {Vidal}(2007)}]{PhysRevLett.99.220405}%
  \BibitemOpen
  \bibfield  {author} {\bibinfo {author} {\bibfnamefont {G.}~\bibnamefont
  {Vidal}},\ }\href {\doibase 10.1103/PhysRevLett.99.220405} {\bibfield
  {journal} {\bibinfo  {journal} {Phys. Rev. Lett.}\ }\textbf {\bibinfo
  {volume} {99}},\ \bibinfo {pages} {220405} (\bibinfo {year}
  {2007})}\BibitemShut {NoStop}%
\bibitem [{\citenamefont {Aguado}\ and\ \citenamefont
  {Vidal}(2008)}]{aguado2008entanglement}%
  \BibitemOpen
  \bibfield  {author} {\bibinfo {author} {\bibfnamefont {M.}~\bibnamefont
  {Aguado}}\ and\ \bibinfo {author} {\bibfnamefont {G.}~\bibnamefont {Vidal}},\
  }\href@noop {} {\bibfield  {journal} {\bibinfo  {journal} {Physical review
  letters}\ }\textbf {\bibinfo {volume} {100}},\ \bibinfo {pages} {070404}
  (\bibinfo {year} {2008})}\BibitemShut {NoStop}%
\end{thebibliography}%

\end{document}